\documentclass[%
 reprint,
superscriptaddress,
nofootinbib,
 amsmath,amssymb,
 aps,
prd,
]{revtex4-2}

\usepackage{graphicx}%
\usepackage{dcolumn}%
\usepackage{bm}%
\usepackage{siunitx}
\usepackage{braket}
\usepackage{gensymb}
\usepackage{aas_macros}
\usepackage{subcaption}

\usepackage[usenames, dvipsnames]{xcolor}

\newcommand{\refresp}[1]{#1}

\newcommand{\vect}[1]{\ensuremath{\mathbf{#1}}}

\newcommand{\likelihood}{\mathcal{L}}
\newcommand{\paramvol}{\ensuremath{\mathrm{d}\theta}}
\newcommand{\evidence}{\mathcal{Z}}
\newcommand{\code}[1]{\texttt{#1}}
\newcommand{\bayesf}{\ensuremath{\mathcal{B}}}

\newcommand{\Msun}{\ensuremath{M_\odot}}

\newcommand{\kepler}{\textit{Kepler}}
\newcommand{\keplerfullname}{\textit{Kepler Space Telescope}}
\newcommand{\keplerfov}{105 $\unit{\deg^2}$}
\newcommand{\keplernstars}{\num{1.6e5}}

\newcommand{\romanname}{\textit{Roman}}
\newcommand{\romanfullname}{\textit{Nancy Grace Roman Space Telescope}}
\newcommand{\romanfov}{0.281 $\unit{\deg^2}$}
\newcommand{\romannstars}{10^8}

\newcommand{\subsetnnstars}{10^3}

\newcommand{\freq}{f_I}
\newcommand{\logfreq}{\log_{10} \freq}
\newcommand{\strain}{h_0}
\newcommand{\logstrain}{\log_{10} \strain}
\newcommand{\cosinc}{\cos i}
\newcommand{\positionangles}{(\phi, \theta)}
\newcommand{\polarizationangle}{\psi}
\newcommand{\initphase}{\Phi_0}
\newcommand{\lumdist}{d_L}
\newcommand{\chirpmass}{\mathcal{M}_c}

\newcommand{\kepleravgstrainlim}{\ensuremath{10^{-12.45}}}
\newcommand{\romanavgstrainlim}{\ensuremath{10^{-11.39}}}
\newcommand{\kepleravglumdist}{3.6 Mpc}
\newcommand{\romanavglumdist}{0.31 Mpc}

\begin{document}

\preprint{APS/123-QED}

\title{A Fast Bayesian Method for Coherent Gravitational Wave Searches with Relative Astrometry}%

\author{Benjamin Zhang}
\email{zhangben@usc.edu}
\affiliation{University of Southern California, Los Angeles, CA 90089, USA}
\author{Kris Pardo}
\affiliation{University of Southern California, Los Angeles, CA 90089, USA}
\author{Yijun Wang}
\affiliation{California Institute of Technology, Pasadena, CA 91125, USA}
\author{Luke Bouma}
\affiliation{Carnegie Observatories, Pasadena, CA 91101, USA}
\author{Tzu-Ching Chang}
\author{Olivier Doré}
\affiliation{California Institute of Technology, Pasadena, CA 91125, USA}
\affiliation{Jet Propulsion Laboratory, California Institute of Technology, Pasadena, CA 91101, USA}

\date{\today}%

\begin{abstract}
    Using relative stellar astrometry for the detection of coherent gravitational wave sources is a promising method for the microhertz range, where no dedicated detectors currently exist.
    Compared to other gravitational wave detection techniques, astrometry operates in an extreme high-baseline-number and low-SNR-per-baseline limit, which leads to computational difficulties when using conventional Bayesian search techniques.
    We extend a technique for efficiently searching pulsar timing array datasets through the precomputation of inner products in the Bayesian likelihood, showing that it is applicable to astrometric datasets.
    Using this technique, we are able to reduce the total dataset size by up to a factor of $\mathcal{O}(100)$, while remaining accurate to within 1\% over two orders of magnitude in gravitational wave frequency.
    Applying this technique to simulated astrometric datasets for the \textit{Kepler Space Telescope} and \textit{Nancy Grace Roman Space Telescope} missions, we obtain forecasts for the sensitivity of these missions to coherent gravitational waves.
    Due to the low angular sky coverage of astrometric baselines, we find that coherent gravitational wave sources are poorly localized on the sky.
    Despite this, from $10^{-8}$ Hz to $10^{-6}$ Hz, we find that \textit{Roman} is sensitive to coherent gravitational waves with an instantaneous strain above $h_0 \simeq 10^{-11.4}$, and \textit{Kepler} is sensitive to strains above $h_0 \simeq $ $10^{-12.4}$.
    At this strain, we can detect a source with a frequency of $10^{-7}$ Hz and a chirp mass of $10^9$ $M_\odot$ at a luminosity distance of 3.6 Mpc for \textit{Kepler}, and 0.3 Mpc for \textit{Roman}.
    We finally discuss possible strategies for improving on these strain thresholds.
\end{abstract}

\maketitle

\section{Introduction}

    Gravitational waves from compact binary mergers of neutron stars and black holes were first directly detected by the Laser Interferometer Gravitational-Wave Observatory (LIGO) \cite{GW150914}.
    Following this detection, evidence for a stochastic gravitational wave background (GWB) from the merger of supermassive black hole binaries (SMBHBs) has recently been found by the NANOGrav pulsar timing array (PTA) \cite{agazie+23_nanograv_15yr_gwb}, and subsequently by other PTAs \cite{eptacollaborationandinptacollaboration:_2023_SecondDataRelease, xu_2023_SearchingNanoHertzStochastic, reardon_2023_SearchIsotropicGravitationalwave}.
    As the SMBHB GWB may lead to beyond-Standard Model physics \cite{afzal+2023_nanograv_15yr_bsm}, constraining it is important.
    One source that can have an outsize effect on the fitted parameters of the background is the presence of coherent GW signals from individual SMBHB sources.
    As the background is composed of these discrete sources, contributions from nearby sources can significantly ``spike" the power of the background at a specific frequency \cite{agazie_2024_NANOGrav15Yr_discreteGWB}.

    In contrast to the LIGO frequency band of $10^{1}$-$10^{3}$ $\unit{Hz}$, the GW frequency band probed by PTAs is roughly $10^{-9}$-$10^{-7}$ $\unit{Hz}$.
    Sensitivity in this range is bounded by the total pulsar observation time on the lower end, and by noise from pulsar spindown fitting on the upper end.
    Within this band, NANOGrav has not yet conclusively detected coherent GW sources, though analysis of the background suggests that as-yet unresolved sources may be present \cite{agazie_2023_NANOGrav15yearData_coherentsearch, agazie_2024_NANOGrav15Yr_discreteGWB}.
    Once detected, coherent GW sources in the PTA frequency range are expected to be from the slow inspiral phase of SMBHBs with chirp masses of $10^7$ to $10^{10}$ $\Msun$.
    The upcoming LISA mission is also expected to detect coherent GWs from the mergers of intermediate-mass black holes, in between $10^{-5}$ and $10^{-1}$ $\unit{Hz}$ \cite{amaro-seoane_2017_LaserInterferometerSpace, colpi_2024_LISADefinitionStudy}.
    A frequency gap in GW detection therefore currently exists between $10^{-7}$ and $10^{-5}$ $\unit{Hz}$, the ``microhertz regime", in which we expect to see GW signals from more rapidly-upchirping SMBHBs in the low frequency end, and their final mergers in the high-frequency end.
    Besides astrophysical binaries, signals of cosmological interest may also exist in this range, such as from cosmic strings and primordial black hole production mechanisms \cite{chang_2020_StochasticGravitationalWave, domenech_2024_ProbingPrimordialBlack}.

    Detection of gravitational waves using astrometry offers a promising method for covering this microhertz frequency gap.
    Just as gravitational waves induce effective path length changes for incoming radio-frequency light, which manifest as timing residuals for PTAs, GWs passing by Earth also induce transverse light deflections \cite{pyne_1996_GravitationalRadiationVerya, book_2011_AstrometricEffectsStochastica}.
    These transverse deflections manifest as changes in the angular positions of stars, and can be detected with sufficiently accurate astrometry using space-based photometric surveys.
    Missions such as GAIA that provide absolute astrometry have been forecast to be more sensitive than PTAs at frequencies greater than $10^{-7.5}$ $\unit{Hz}$ \cite{klioner_2018_GaialikeAstrometryGravitational, moore_2017_gaia_compression}. However, specific raw data products, which have not been released by the GAIA team, are needed to properly search for GWs in the higher frequencies. Even with relative astrometry instead of absolute, the astrometry method promises better sensitivity than PTAs for frequencies above $10^{-7}$ $\unit{Hz}$ using the upcoming \romanfullname{} \cite{wang+21_romangw_coherent, wang+22_romangw_background, pardo_2023_GravitationalWaveDetectiona}.

    Computationally, conducting a coherent GW search using astrometric data presents unique challenges.
    Generally, the signal-to-noise per stellar baseline for coherent GW signals is expected to be extremely low, on the order of $10^{-6}$.
    This is much lower than the time-integrated SNR per baseline of $\sim 1$ for dedicated instrumental detectors such as LIGO/VIRGO, or $\sim 10^{-2}$ for PTAs.
    While the low SNR per baseline for astrometric detection is compensated for by observing many baselines ($10^5$ -- $10^8$) with a short observing cadence (15 -- 30 min), this leads to dataset sizes at least on the order of a terabyte.
    Conventional Bayesian methods such as MCMC for searching for coherent GWs are therefore computationally infeasible.
    The dataset also cannot be searched using a shorter sliding time window as LIGO-like detectors do, as coherent gravitational waves at the low-frequency end are expected to have cycles that span the entire observing length.
    Averaging the astrometric deflections of clusters of stellar baselines has been proposed for reducing the size of the GAIA astrometric dataset, and in principle results in a $<1\%$ reduction in sensitivity \cite{moore_2017_gaia_compression}.
    However, this may exacerbate spatially-correlated systematics in the data.
    In addition, this technique may not work as well for surveys that only cover a small fraction of the sky.

    For PTAs, Ref.~\cite{becsy+22_quickcw1} has developed a technique for speeding up Bayesian likelihood evaluations of a coherent GW model on pulsar timing residual data.
    This technique exploits the approximately constant-frequency sinusoidal structure of coherent GWs in the nanohertz limit, and precomputes inner products between these sinusoids and the timing data.
    The Bayesian likelihood can then be exactly rewritten as a linear combination of these inner products.
    Due to this precomputation, for a given GW frequency, this technique also reduces the total dataset size.
    Ref.~\cite{becsy+24_quickcw2} extends this technique to work with variable GW frequencies, through the use of cubic interpolation of the precomputed inner products across a linearly-spaced grid in GW frequency.

    In this work, we generalize this dataset size reduction technique to operate on astrometric deflections, and find that the Bayesian likelihoods it produces remain accurate across 2 orders of magnitude in frequency, even when using less-accurate linear interpolation across a logarithmically-spaced frequency grid.
    Using this technique, we can reduce dataset size to a manageable amount while avoiding averaging together baselines.\footnote{Although both techniques are complementary, and so can be combined if needed.}
    We then provide a coherent GW sensitivity forecast for \romanname{}  using realistic mock data, and in addition provide a forecast for archival photometric data from the \keplerfullname{}.

    In Section \ref{sec:methods}, we review the theoretical basis of GW detection using astrometry and the survey characteristics of \romanname{}  and \kepler{}.
    We detail our data reduction method, and the Bayesian search pipeline we use for detecting coherent GWs.
    We compute the accuracy of the data reduction method and provide sensitivity forecasts for \romanname{} and \kepler{} in Section \ref{sec:results}.

    Throughout this work, we adopt the Planck 2018 concordance $\Lambda$CDM cosmology \cite{planckcollaboration_2020_Planck2018Results}.
    
\section{Methods}
\label{sec:methods}

    \subsection{Effect of a coherent GW on stellar astrometry}
    \label{sec:methods:astrometry}

        \begin{figure*}
            \centering
            \includegraphics[width=0.45\textwidth]{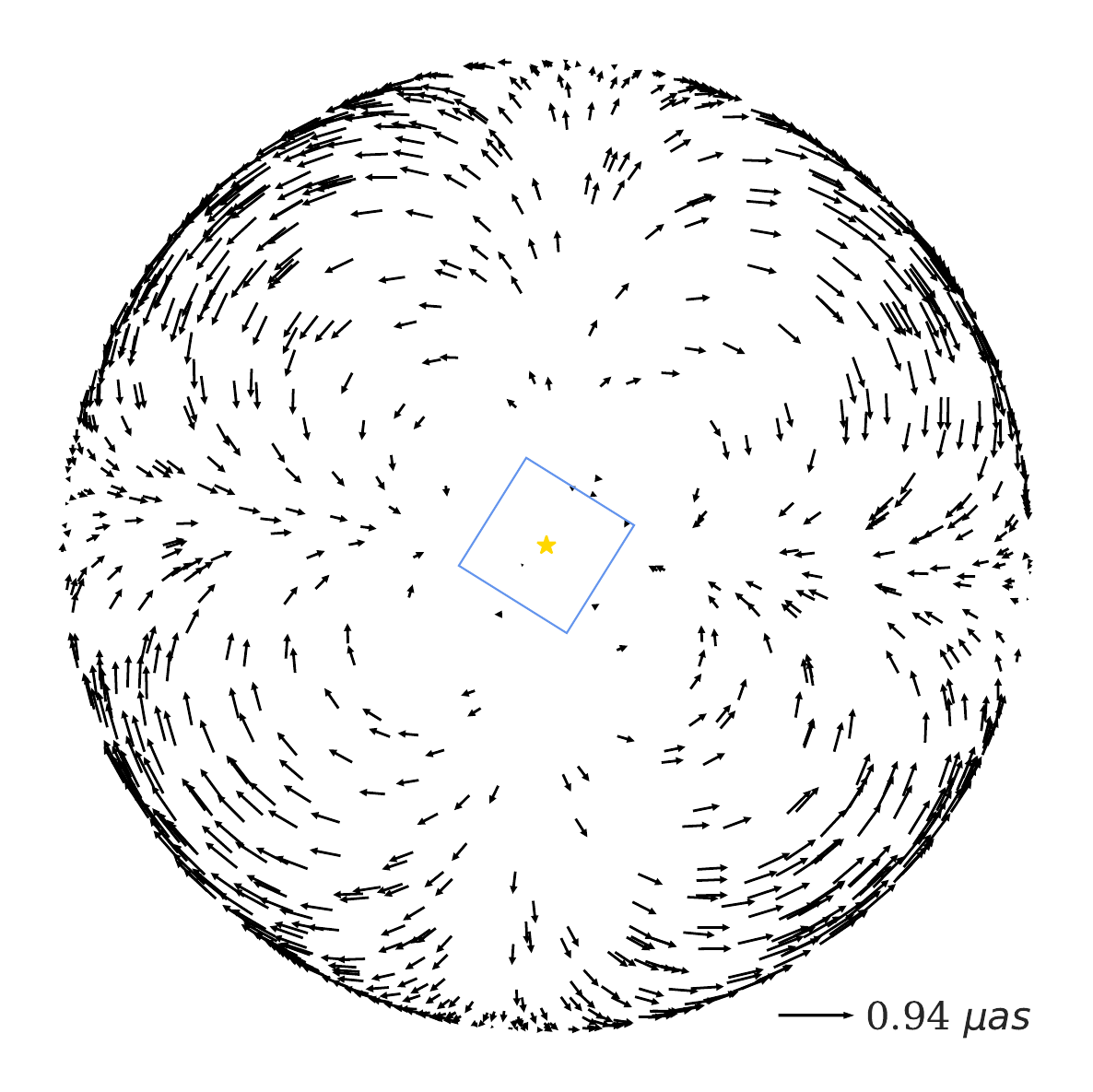}
            \includegraphics[width=0.45\textwidth]{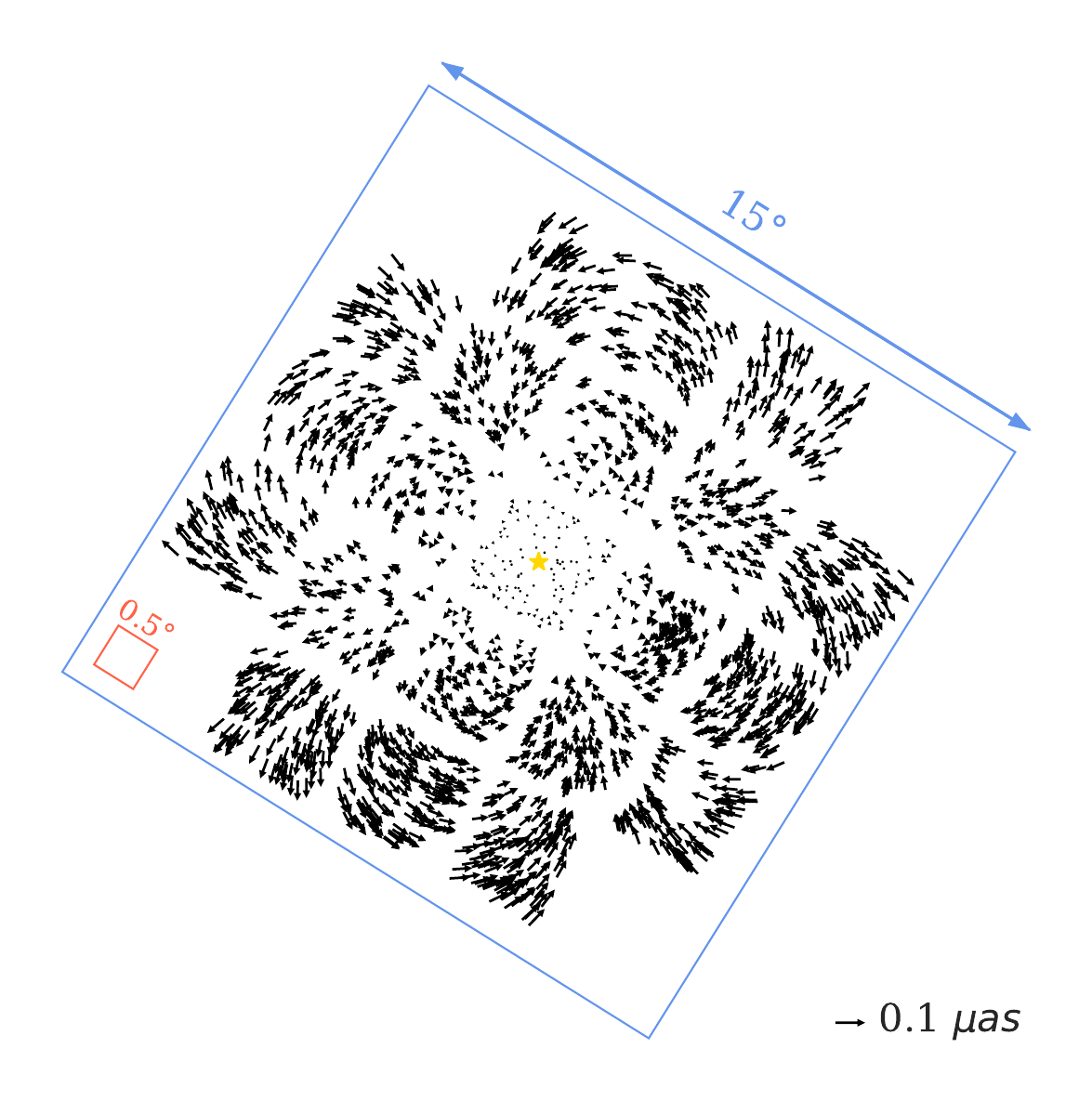}
            \caption{Astrometric deflections from a single coherent GW source at the position of the gold star, with a chirp mass of $\chirpmass{} = 10^9$ $\Msun{}$ and a frequency of $\freq{} = 10^{-6}$ $\unit{Hz}$. 
            The tail of each arrow represents a star's position if no GW is present, while the tip represents its apparent position with the coherent GW present.
            The left subplot shows the deflections for 1000 example stars over one hemisphere of the sky, reproduced from Ref.~\cite{wang+21_romangw_coherent}.
            The right subplot shows the Kepler field of view with deflections for 2000 example stars.
            The blue box of side length $15\degree$ is present on both subplots to get an approximate sense of scale for \kepler{}'s footprint, but does not indicate its actual maximum angular extent.
            The actual pattern of \kepler{}'s 105 $\unit{deg}^2$ FoV can be seen from the example stars.
            The field of view of \romanname{} is also shown for scale as the red box on the right subplot.}
            \label{fig:example-gw-deflection}
        \end{figure*}

        The angular response of measured stellar centroids (hereafter ``deflections") to a coherent GW passing through Earth forms a quadrupolar pattern, with the deflection for each star tracing out an ellipse with time.
        We outline the theoretical response of measured stellar centroids to a coherent GW here, briefly summarizing the exact deflection response derived by Ref.~\cite{book_2011_AstrometricEffectsStochastica}.

        We assume the distant-source limit, where the luminosity distance to the GW source $\lumdist$ is much greater than the GW wavelength $\lambda_{GW}$.
        Under this limit, the total effect of a GW only comes from its effect at every star (the ``source term") and its effect at Earth (the ``observer term").
        Since the stars we observe are at different distances from any GW source, the deflection of each star from its source term will be uncorrelated from star to star.
        We therefore neglect the source term for our analysis, treating it as additional stochastic noise.
        With these assumptions, the angular deflections $dn$ for a star at sky position $\vect{n}$ generated by a coherent GW source at a sky position $\vect{q}$ are
        \begin{equation}
        \begin{split}
            \label{eqn:dn}
            dn^{i}(t,\vect{n}) = \frac{n^i-q^i}{2\left(1-\vect{q}\cdot\vect{n}\right)}&h_{jk}(t,\vect{0})n^jn^k\\&-\frac{1}{2}h^{ij}(t,\vect{0})n_j
        \end{split}\,,
        \end{equation}
        where $h_{ij}(t, \vect{0})$ is the general GW strain tensor evaluated at the observer's position $\vect{0}$.

        We additionally assume the Newtonian ``0PN" circular binary approximation for the GW, where the $h$-tensor takes the form of a plane wave with time and source-dependent frequency $f(\chirpmass{}, \freq{}, t)$
        \begin{equation}
            h_{ij}(t,\vect{x}) =\strain {\rm{Re}}\left[H_{ij}e^{-i2\pi \left(t+\vect{q}\cdot\vect{x}\right) f(\chirpmass, f_i, t) + i2\pi \initphase} \right] \,,
        \end{equation}
        where $\chirpmass$ is the chirp mass, and $\freq$ is the initial frequency at $t = 0$, the start of observations.
        $\initphase$ is an arbitrary global phase, associated with the time we start observing the GW source.

        $\strain$ is defined as the instantaneous GW strain, a combination of the frequency $f$, the chirp mass $\chirpmass$, and the luminosity distance to source $\lumdist$:
        \begin{equation}
            \strain = \frac{2G^{5/3}}{c^4}(\pi f)^{2/3}\frac{\chirpmass^{5/3}}{\lumdist}\,.
        \end{equation}
        If we assume that the GW source is a circular binary, $H_{ij}$ is the polarization basis tensor, which depends on the binary orbital inclination $i$, the position angles of the source on the sky $\positionangles$, and the polarization angle $\polarizationangle$. 
        $\polarizationangle$ is defined as the angle between the major axis of the sky-projected binary orbital ellipse, and the equatorial plane of the galactic $(l, b)$ coordinate system.
        In combination, $\phi$, $\theta$ and $\polarizationangle$ describe the full $\text{SO}(3)$ rotation of a binary source relative to the observer.
        The luminosity distance $\lumdist$ can be determined from $\strain$, $f$, and $\chirpmass$, and so is not a free parameter.
        The free coherent GW model parameters are summarized in Table \ref{tab:model-priors}.

    \subsection{Astrometric survey data}
    \label{sec:methods:data}

        In this work, we aim to forecast coherent GW detection prospects for two missions, the \romanfullname{} and \keplerfullname{} (hereafter \romanname{} and \kepler{}). 
        In this section, we describe the characteristics of each survey, and our simulated datasets for the forecast. 

        We focus on the Galactic Bulge Time Domain Survey of \romanname{}, which is primarily aimed at detecting exoplanets through the microlensing method \cite{gaudi_2023_RomanGalacticExoplanet}. Although the detailed survey strategy and parameters are still to be determined, we use the ones proposed in Ref.~\cite{gaudi_2019_AuxiliaryScienceWFIRST}.
        With this strategy, \romanname{}  would take exposures of $\romannstars$ stars in the Milky Way bulge across 7 fields of size $\sim$ \romanfov{}, with a cadence of 15 min between exposures.
        With an expected astrometric precision of 1.1 $\unit{mas}$ per exposure and total observing time of six 72-day seasons, \romanname{}  would be in principle an ideal mission to perform astrometric GW detection.
        However, the detailed pointing control of \romanname{} can drastically affect the prospects of GW detection via astrometry, due to ``mean-signal subtraction".
        The reaction wheels of \romanname{} are constantly engaged so that the telescope is centered with respect to guide stars within the field of view.
        Because these guide stars also experience deflections from a coherent GW, only deflections relative to the mean across the field of view are measurable.
        This is projected to decrease sensitivity by a factor of $\sim 100$ compared to the scenario where \romanname{} observes each field in free fall for the entire observing season \cite{wang+21_romangw_coherent, pardo_2023_GravitationalWaveDetectiona}.

        We also assess the astrometric GW sensitivity of the primary \kepler{} mission
        \cite{Borucki2010}, which ran from May 2009 to August 2013.  While \kepler{}
        was designed to detect exoplanet transits, it has many desirable properties for
        astrometric GW detection.  Over sixteen 93-day quarters, \kepler{} observed the
        same \keplerfov{} field with a 30-minute cadence.\footnote{While Kepler
        observed over 18 quarters in total, the first two quarters had poor astrometric
        performance due to variable stars being present in the list of fine guidance
        stars \cite{VanCleve2016}.  We discard those quarters from consideration.} Each
        exposure contains $\sim$$\keplernstars$ stars, with a median astrometric
        precision per exposure of 0.7 $\unit{mas}$
        \cite[e.g.][]{monet_2010_PreliminaryAstrometricResults}.
        While it is currently unclear if the pointing control of \kepler{} results in the same mean-signal subtraction as \romanname{}, if mean-signal subtraction is present, the larger field of view of \kepler{} will result in a smaller sensitivity decrease than \romanname{}.
        For the worst case where the mean GW signal is subtracted from every exposure, the \kepler{} sensitivity is projected from simulations to decrease by a factor of $\sim$ 10.
        As the degree of mean-subtraction is currently unknown, for this work we optimistically assume that \kepler{} data is unaffected by mean-subtraction.
        
        We show an example of astrometric deflections from a coherent GW within the \kepler{} field of view in Figure~\ref{fig:example-gw-deflection}, with one sky hemisphere and \romanname{}'s field of view for scale.

        Data from \kepler{} and \romanname{} must be further reduced before conducting a coherent GW search, due to the presence of various systematic effects.
        These systematics generally produce orders-of-magnitude greater deflection than any expected coherent GW signal.
        Some systematics include differential velocity aberration, radiation pressure, thermal expansion of the telescope's focal plane and optics, and others.
        For the remainder of this work, we optimistically assume that these systematics are either fully removed, or that the frequency ranges they occupy are excluded from any coherent GW search.
        \refresp{Although in reality systematics will likely be impossible to fully remove, if periodic they are expected to affect narrow frequency ranges, or if aperiodic, be uncorrelated across observed stars. For example, the differential velocity aberration (DVA) effect is a coherent apparent expansion or contraction of the observed star field which arises from the orbital motion of \kepler{}.
        DVA is expected to be a major systematic, on the order of 400-2000 $\unit{mas}$ \citep{VanCleve2016}; however, as it comes from \kepler{}'s orbit, imperfect cleaning will in effect only result in apparent stellar motion at a frequency of $f \simeq 1 / \unit{yr}$.
        The sensitivity to coherent GWs will therefore be decreased only at narrow frequency bands centered around $1 / \unit{yr}$ and its harmonics.
        For an example of an aperiodic systematic, the median proper motion of Kepler stars as measured by GAIA is 1.09 $\unit{mas}/\unit{yr}$ \citep{brown_2011_kic, gaia_2023_dr3}. 
        Since the proper motion is linear, it can be fit out on a star-by-star basis.
        Since the proper motion is uncorrelated between stars \cite{janes_2017_kepler_commonpropermotions}, imperfect fits will only result in additional stochastic noise when taken across the entire dataset.}

        For the mock \kepler{} and \romanname{} survey data we use in the remainder of this paper for forecasting, we assume white astrometric noise.
        For \kepler, we use the \kepler{} magnitudes $K_p$ of stars from the archival data to assign an astrometric noise level to each star from a pre-calibrated curve.
        This curve is fit from real data on \kepler's astrometric performance \cite{monet_2010_PreliminaryAstrometricResults}.
        The median astrometric noise per exposure is $\sim 0.74$ $\unit{mas}$.
        For \romanname{}, we assume uniform astrometric noise across all stars as an approximation (though in reality, astrometric noise will depend on stellar magnitude like \kepler).
        We assign all stars an astrometric per-exposure noise of 1.1 $\unit{mas}$ \cite{wfirst+19_roman_astrometry}.
        As previously mentioned, we optimistically assume full removal of systematic effects, and so do not add them to our mock survey data.
        Finally, for both surveys, we assume the flat-sky approximation, fixing the tangent vectors in Equation \ref{eqn:dn} for all stars within the field of view to the tangent vectors in the field of view center.

    \subsection{Inner product precomputation}
    \label{sec:methods:inner-prod}

        The high number of stellar baselines involved in astrometric GW detection combined with a high observing cadence presents a unique big data challenge.
        For example, \kepler, assuming double-precision stellar centroid measurements along two flat-sky axes, has an estimated dataset size of:
        \begin{equation*}
            \keplernstars \text{ stars} \times \frac{16 \times \qty{93}{days}}{\qty{30}{min}} \times \qty{2}{axes} \times \qty{64}{bits} = \qty{183}{GB}
        \end{equation*}
        Along these lines, the \romanname{}  dataset is estimated to be around $\qty{66}{TB}$.
        This also assumes the most optimistic noise characterization, where the centroid noise for each star is described by a single Gaussian variance, which remains constant over the entire observation period.
        For a more pessimistic scenario, we consider a case where correlated noise is present in the stellar centroids below a certain timescale after data cleaning, necessitating the storage of a banded covariance matrix.
        Here as an example we take this timescale to be the time between angular momentum dumps (where spacecraft thrusters are used to rid the station-keeping reaction wheels of their accumulated angular momentum).
        This is every 3 days for \kepler{}, and every $\sim38$ hours for \romanname{}.
        Under this assumption, the expected dataset size swells to $\sim \qty{13}{TB}$ for \kepler{}, and $\qty{5}{PB}$ for \romanname{}.
        Even on the most optimistic end, datasets of this size cannot fully fit into random-access memory.
        Naively then, any likelihood evaluations used in a Bayesian search will be limited by hard drive access speed, making Bayesian techniques such as nested sampling or MCMC prohibitively computationally expensive.

        We therefore seek a method of reducing the size of the astrometric dataset before performing likelihood evaluations.
        If we assume that any candidate coherent GW signal 
        has no frequency evolution (i.e. the $h$-tensor is $\propto \cos{(\omega t)}$ for some non-changing GW frequency $\omega$), then this can be accomplished by reformulating the Bayesian likelihood.

        This reformulation was first developed by Ref.~\cite{becsy+22_quickcw1} for reducing the computational cost of evaluating PTA likelihoods, which we briefly summarize here.
        Ref.~\cite{becsy+22_quickcw1} gives a general treatment for slowly-evolving GWs in the stationary-phase approximation; however, for our purposes we assume that the frequency and amplitude of the GW are approximately constant over the observation window.\footnote{As an example of a source we might observe, the ``chirp time" $\tau_{\text{chirp}} \equiv f / \dot{f}$ for a binary with chirp mass $\chirpmass{} = 10^9 \Msun{}$ and initial frequency $\freq{} = 10^{-7} \unit{Hz}$ is 40 years.}
        As stated above, we assume that the GW frequency at Earth $\omega$ remains constant over the observation time.
        For pulsar timing, the ``pulsar term" frequency $\omega_p$ is also important.
        This is the frequency of the GW when it arrives at a pulsar.
        In almost all cases, $\omega \neq \omega_p$, as the distance between the GW source and the pulsar will be significantly different from the source-observer distance.
        While the GW frequency stays approximately constant over the length of observations (1-10 yr), frequency evolution still means that the GW frequency at the pulsar and the observer will significantly differ.
        Considering only one pulsar for simplicity, the pulsar timing residual signal $s(t, \tilde{\theta})$ can be expressed as a linear combination of filter functions $S^i(\omega, \omega_p, t)$: 
        \begin{align*}
            S^1(\omega, t) &= \cos{\omega t} \\
            S^2(\omega, t) &= \sin{\omega t} \\
            S^3(\omega_p, t) &= \cos{\omega_p t} \\
            S^4(\omega_p, t) &= \sin{\omega_p t}
        \end{align*}
        We define the shorthand $\braket{\vect{a}|\vect{b}} \equiv \vect{a}^T C^{-1} \vect{b}$ for some arbitrary time-series $\vect{a}, \vect{b}$ and data covariance matrix $C$.
        The pulsar timing residual log-likelihood $\log \mathcal{L} \equiv -\frac{1}{2} \braket{\delta \vect{t} - \vect{s}|\delta \vect{t} - \vect{s}}$ can then be exactly rewritten as a linear combination of the inner products $\braket{\delta \vect{t} | \vect{S}^i}$ and $\braket{\vect{S}^j | \vect{S}^k}$.
        If these inner products are pre-computed once, all variation in non-frequency-related GW model parameters is entirely contained within the linear combination coefficients, drastically reducing the computational cost of likelihood evaluations.

        For a range of candidate GW frequencies, Ref.~\cite{becsy+24_quickcw2} further extend this scheme by evaluating the inner products at a linearly-spaced grid of candidate frequencies, and performing cubic approximation of the inner products between grid points.
        With a frequency grid spacing of 10 points per $1 / T_{\text{obs}}$ sized frequency bin (where $T_{\text{obs}}$ is the total observation time), they find a $< 1 \%$ accuracy loss in the likelihood (with frequency evolution included).

        We find that this technique is also directly applicable to astrometric GW deflections.
        As we treat the effect of the source term as stochastic noise due to the vastly larger number of baselines we observe compared to PTAs, our only two filter functions are $\cos \omega \vect{t}$ and $\sin \omega \vect{t}$.
        For deflections along two arbitrarily-chosen axes $\vect{d}_x$ and $\vect{d}_y$, the log-likelihood can still be rewritten as a linear combination of inner products.
        We split the full covariance $C$ into three unique blocks $C_{xx}, C_{xy}, C_{yy}$, where for example $C_{xx}$ is the covariance between the x-axis deflection and itself.
        The inner products we precompute are then $\braket{\vect{d}_a | C_{ab}^{-1} | \cos \omega \vect{t}}$, $\braket{\vect{d}_a | C_{ab}^{-1} | \sin \omega \vect{t}}$, and $\braket{\cos \omega \vect{t} | C_{ab}^{-1} | \sin \omega \vect{t}}$.
        For a given GW frequency $\freq$, precomputing these inner products reduces our dataset size from $\mathcal{O}(NT)$ to $\mathcal{O}(N)$, where $N$ is the number of stellar baselines and $T$ the number of astrometric exposures.

        In order to let $\freq$ vary, we compute these inner products on a frequency grid like Ref.~\cite{becsy+24_quickcw2} and interpolate between them.
        However, rather than choosing grid points evenly spaced in $\freq$ with cubic interpolation, we choose a grid evenly spaced in $\logfreq$ over our frequency prior, and perform linear interpolation for performance reasons.
        We opt for an evenly-spaced grid in $\logfreq$ due to the wider frequency range we search compared to PTAs.
        Ref.~\cite{becsy+24_quickcw2} evaluate the interpolation accuracy for a frequency prior of 0-32 $\unit{nHz}$; in contrast, we search over two orders of magnitude from 10 $\unit{nHz}$-1 $\unit{\mu Hz}$.
        Future searches will likely extend another two orders of magnitude, up to $\sim 10^{-4}$ $\unit{Hz}$.
        However, as we assume constant $\freq{}$ over our observing span, we omit the $10^{-6}-10^{-4}$ $\unit{Hz}$ frequency span from our current analysis, as sources in this range are expected to have a chirp time $\tau_{\text{chirp}} \equiv f / \dot{f}$ shorter than our total observation time, and so will be significantly affected by GW frequency evolution.

        With $F$ grid points in $\logfreq$, doing this interpolation increases the compressed dataset size to $\mathcal{O}(NF)$, but still provides savings compared to the raw data if $F < T$.
        In Section \ref{sec:results:inner-prod}, we compute the $F$ necessary to approximate the Bayesian evidence $\evidence{}$ to 1\% or better for a barely-detectable GW source.

    \subsection{Bayesian coherent GW search}
    \label{sec:methods:bayesian-search}

        We conduct our search for coherent GWs using the Bayesian nested sampling \cite{skilling06_nestedsampling} method to sample posteriors for the coherent GW parameters described in Section \ref{sec:methods:astrometry}.
        Nested sampling obtains the Bayesian evidence $\evidence$ by sampling from a portion of the Bayesian prior space above a continually rising likelihood threshold.
        For nearly all well-behaved likelihoods of interest, this portion of prior space exponentially shrinks as the sampling progresses.
        The convergence of nested sampling is naturally determined by the marginal evidence per new sample falling below a threshold.
        After sampling completes, the Bayesian posterior can be obtained by through piecewise integration using the sampled points.
        
        Our choice of nested sampling is motivated by the small field of view of astrometric surveys.
        This small field of view results in a poor ability to localize coherent GWs, which then causes multi-modal likelihoods.
        As a toy example, consider a coherent GW source which is 90$\degree$ from the center of the telescope's field of view.
        Due to the small field of view, the deflections caused by this source appear almost uniform across the entire field, so that the initial GW phase $\initphase$ is approximately degenerate with the exact sky location along the great circle which is 90$\degree$ from the field of view.
        In addition, for a source with a face-on inclination (so that the binary orbit is also face-on), changing the polarization angle $\polarizationangle$ is exactly equivalent to changing the initial GW phase $\initphase$.
        This leads to a multi-modal likelihood for these two parameters.
        Nested sampling naturally handles these multi-modal likelihoods, due to the non-local nature of the sampling process: for each sampling step, all likelihood peaks above the threshold can (ideally) be sampled.
        In addition, nested sampling yields the Bayesian evidence $\evidence$, which we use for assessing the relative strength of the null ``no GW signal present" hypothesis.
        
        We use the Gaussian log-likelihood described in Section \ref{sec:methods:inner-prod}, which is reformulated in terms of data-signal inner products.
        We note again that this likelihood omits any ``star term" coherent GW response at the star's position, implicitly treating it as additional stochastic noise. 

        We implement the likelihood using the \code{JAX} library \cite{jaxgithub}, and conduct nested sampling using the \code{jaxns} code \cite{albert20_jaxns1, albert23_jaxns2}.
        For reproducibility, we detail our \code{jaxns} version and precise tuning in Appendix \ref{appendix:nested-sampling}.

        Our priors are summarized in Table \ref{tab:model-priors}. 
        We adopt linearly uniform priors for the cosine of the orbital inclination $\cos{i}$, the orbital polarization angle $\polarizationangle$, and the global initial GW phase $\initphase$.
        We adopt a uniform prior across the celestial sphere for the source position, which for the source position angles $\positionangles$ implies a uniform prior for $\phi$ and a uniform prior for $\cos{\theta}$.
        Internally, these angles are for the spherical coordinate system which has its z-axis aligned with the telescope FoV center. The coordinate system's x-axis is chosen to be parallel with the galactic coordinate system's equatorial plane ($b = 0$), and its y-axis is chosen to form a right-handed coordinate system.
        We also choose to use a linearly uniform prior for the instantaneous GW strain $\strain$, rather than a prior which is uniform in $\logstrain$.
        Due to the efficiency of nested sampling, we find that an $\strain$-uniform prior still converges well for injected strains which are orders of magnitude less than the $\strain$ prior upper limit.
        Finally, we adopt a $\log_{10}$-uniform prior for the GW frequency $\freq{}$.

        \begin{table}[]
            \centering
            \begin{tabular}{|l|l|}
            \hline
            GW model parameter                  & Uniform prior limits                                           \\ \hline
            GW frequency $\logfreq$ [$\log_{10} (f_I / \unit{Hz})$]          & {[}-8, -6{]}                                                   \\
            Instantaneous GW strain $\strain$       & {[}$10^{-18}$, $10^{-11}${]} \\

            Cosine of source orbital inclination $\cosinc$ & {[}-1, 1{]} \\
            
            Source position angle $\phi$        & {[}0, $2 \pi$)                                                 \\
            Source position angle $\cos \theta$ & {[}-1, 1{]}                                                    \\
            Polarization angle $\polarizationangle$           & {[}0, $\pi$)                                                      \\
            Initial GW phase $\initphase$           & {[}0, $2 \pi$)                                                 \\ \hline
            \end{tabular}
            \caption{Bayesian prior limits for our coherent GW model. While the physically correct lower limit for a uniform $\strain$ prior is 0, we choose $10^{-18}$ instead for numerical stability reasons. As with the PTA timing residual response, subtracting $\pi$ from $\polarizationangle$ produces an equivalent deflection response to adding $\pi$ to $\initphase$, and so the upper prior limit of $\polarizationangle$ can be reduced to $\pi$.}
            \label{tab:model-priors}
        \end{table}

    \subsection{Model selection}

        As previously mentioned, nested sampling also enables calculating the Bayesian evidence of the coherent GW model, $\evidence_{GW}$.
        We determine our sensitivity to different coherent GW signals by computing the Bayes factor $\bayesf{}$ between $\evidence_{GW}$ and the Bayesian evidence of no GW signal being present, $\evidence_0$.
        $\evidence_0$ is simply the likelihood evaluated at $\strain = 0$.
        Though subjective, $\bayesf{} > 10$ is often taken as ``strong evidence" against the null hypothesis. \cite{kass+95_bayesfactortable}.

        We additionally run into a complication with model selection with the Bayes factor due to our choice of a linearly-uniform prior for $\strain$.
        This prior is chosen to be uniform to be ``uninformative", and indeed as long as the injected strain $\strain$ is within the prior bounds, the posterior is unaffected by our choice of prior limits.
        However, when calculating evidences, ``uninformative" priors are anything but \cite{llorente+22_priorproblems}.
        We consider a case where we choose a prior which is ``too wide" by $A$ orders of magnitude (i.e. the prior upper limit is $10^A$ times the maximum of the recovered posterior).
        While the recovered posterior remains the same as a more restricted prior, the evidence integral contains a factor of the prior density.
        Therefore, the Bayesian evidence of this too-wide prior compared to a prior which tightly encompasses the posterior will be $10^A$ times smaller, potentially leading to false negatives in model selection.

        We circumvent this issue by adopting an approach similar to that of empirical Bayesian prior selection, where the prior extent is a hyperparameter to be optimized over for maximum evidence \cite{petrone+14_empiricalbayes}.
        Because the posterior distribution for $\strain$ is unaffected by the choice of prior limits (assuming that the limits encompass the sampled posterior), we sample $\strain$ with a too-wide prior, and perform a post-hoc correction of the evidence.
        We compute the extent between the 1\% and 99\%iles of the $\strain$ posterior, $10^N$.
        We then multiply $\evidence_{GW}$ by a factor of $10^{W-N}$, where $10^W$ is the extent of the too-wide $\strain$ prior we perform the sampling with.
        This is equivalent to going through the empirical Bayes procedure of maximizing the evidence by shrinking the $\strain$ prior extents, but is more computationally efficient.

        While we also choose an uninformative prior for $\logfreq$, we opt not to perform this evidence correction for it.
        This is because for a log-uniform prior on $f$, changing the extents of the prior would change the sampled posterior of $f$ (in non-log space).

\section{Results \& Discussion}
\label{sec:results}

    \subsection{Inner product precomputation}
    \label{sec:results:inner-prod}

        As described in Section \ref{sec:methods:inner-prod}, we reduce the size of our astrometric dataset by precomputing inner products between the deflection data and sinusoids of different frequencies.
        We compute these inner products for a frequency grid which is evenly spaced in $\logfreq$ over our prior, and linearly interpolate between the frequencies for arbitrary $\logfreq$.
        Here, we give the necessary grid spacing in $\logfreq$ required to approximate the likelihood well.

        In order to assess how well the likelihood is approximated for a given number of frequency grid points over the prior $F$, we can take advantage of the nested sampling method we use.
        One of the results of a nested sampling run is a collection of sampled points from the parameter space, each of which has a likelihood $\likelihood_i$, and an associated parameter-space volume $\paramvol_i$.
        The Bayesian evidence $\evidence{}$ can then be directly obtained by numerical integration: $\evidence{} = \sum \likelihood_i \paramvol_i \equiv \sum \mathrm{d}Z_i$ \cite{skilling06_nestedsampling}.
        In practice, only a subset of points have nontrivial $\mathrm{d} \evidence{}_i$.
        As $\mathrm{d} \evidence{}_i$ is used to weight sampled points when constructing the posterior, we wish to evaluate how well our frequency gridding approximation does for only this subset.

        For both \kepler{} and \romanname{} survey parameters, we start with a nested sampling run where for every nested sampling iteration, we exactly calculate the inner products instead of performing interpolation over a frequency grid.
        Specifically, we consider a nested sampling run on data which has a GW source frequency $\freq{} = 10^{-7} \unit{Hz}$.
        To ensure our interpolation scheme is accurate for marginal-detection cases, we choose the strain of our mock GW source such that the run yields $\bayesf{} = 10$, which is conventionally taken as the threshold for signal detection \cite{kass+95_bayesfactortable}.
        From this run, we obtain the exact Bayesian evidence $\evidence{}_E$, and a collection of sampled points with $\mathrm{d}\evidence{}_{E,i} = \likelihood_{E,i} \paramvol_{E,i}$.

        For some number of frequency interpolation grid points $F$, we construct an approximate likelihood $\likelihood{}_F$ using the interpolation scheme described in Section \ref{sec:methods:inner-prod}.
        Keeping the same $\paramvol_{E,i}$ for each sampled point in the exact nested sampling run, we re-compute the evidence using $\likelihood{}_F$: $\evidence{}(F) = \sum \likelihood{}_{F,i} \paramvol_{E,i}$. \footnote{For likelihoods which are very badly approximated, the assumption that we can reuse the same $\paramvol_{E,i}$ breaks down. This is because later samples in the nested sampling run are causally dependent on the likelihood values of earlier ones, due to the likelihood threshold used to select new samples. An actual nested sampling run with a badly-approximated likelihood will therefore almost certainly diverge significantly from the exact run. However, near the relative evidence error threshold of 1$\%$ we care about, this is not a significant problem.}
        We finally evaluate the relative error in the approximate evidence versus the exact evidence, $|\evidence(F) - \evidence_E| / \evidence_E$.

        \begin{figure*}
            \centering
            \includegraphics[width=0.5\linewidth]{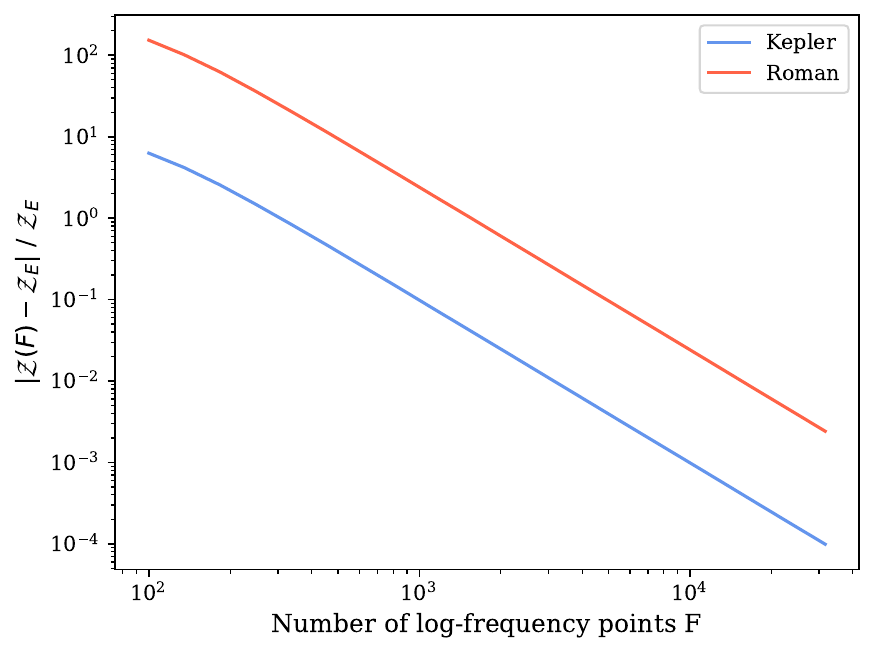}
            \caption{Relative error in the Bayesian evidence $Z(F)$ as a function of the number of points on the GW log-frequency grid used to calculate inner products.
            The relative error is versus the exact evidence $Z_E$, which is the evidence for a nested sampling run where the inner products are evaluated exactly for each sample.
            For this figure, we use single nested sampling runs for \kepler{} and \romanname{}, with a mock GW source frequency $\freq{} = 10^{-7}$ $\unit{Hz}$, and a source strain which yields a Bayes factor $\bayesf{} = 10$.
            Due to the high noise present in the time series for each baseline, the inner products also will be noisy with respect to frequency.
            The linear interpolation in frequency we use for the inner products is therefore unable to perfectly fit the inner product-frequency function even for large values of $F$, which is why the relative error never converges even for $F = 3 \times 10^4$.
            The difference in relative error between the two surveys is solely explained by the number of stars in the data: \keplernstars{} for \kepler{} versus $\romannstars{}$ for \romanname{}.}
            \label{fig:freq-interp}
        \end{figure*}

        The results for this are shown in Figure \ref{fig:freq-interp}.
        For \kepler{}, we need $F = 3140$ in order to approximate the evidence to within $1\%$ of its exact value, and we need $F = 15500$ for \romanname{}.
        If assuming the pessimistic banded covariance case outlined in Section \ref{sec:methods:data}, with this $F$ the \kepler{} dataset size is reduced by a factor of $\sim 360$, and by a factor of $\sim 45$ for \romanname{}.
        The difference in dataset size savings between surveys is solely explained by the number of stars $N_s$ in each survey's dataset.
        As the total log-likelihood consists of a sum of log-likelihoods for each star, we expect the error in $\likelihood{}(F)$ to grow proportionally to $\sqrt{N_s}$, which is exactly reflected in the curves of Figure \ref{fig:freq-interp}.
        Interestingly, the error in the evidence does not appear to depend on either the number of observations per star, or the signal-to-noise of the data.
        This may be because even for large $F$, the linear interpolator is unable to perfectly capture the noisiness in the inner products stemming from the low SNR of the data (reflected in the fact that the evidence error does not converge even for high $F$).
        
        We also find that per sample, the error in the evidence $\mathrm{d} \evidence(F)_i - \mathrm{d} \evidence_{E,i}$ is always negative.
        This suggests that the regions of high likelihood associated with the bulk of the posterior are reflected in rapid changes in the inner product values with respect to frequency, which coarse-gridded likelihoods are less able to capture.

        With these results, we show that the inner product interpolation method from Ref.~\cite{becsy+24_quickcw2} still produces a well-approximated likelihood even when interpolating between points spaced regularly in $\logfreq$.

        We emphasize again that this interpolation method assumes zero frequency evolution of candidate GW waveforms, and so will not work for the high chirp mass/higher frequency end of our observable parameter range, where signals are rapidly upchirping on timescales of months to days.
        We hope to obtain dataset compression while taking frequency evolution into account in a future work.
        
        In addition, we implicitly assume that over many GW cycles, the average position of each star does not drift significantly.
        In essence, rewriting the likelihood in terms of a linear combination of inner products relies on the fact that a sinusoid of arbitrary phase and amplitude can be decomposed into a linear combination of plain sinusoids.
        This is only possible for sinusoids centered at zero, which would no longer be true if 
        a star's position drifts significantly.
        Due to proper motion and non-fully-cleaned systematics, it is likely that stars in our final analysis will exhibit this drift.
        Due to the large number of stars we consider, proper motion will likely not contribute any bias to the inner product-based likelihood; however, remaining systematic effects that cause apparent drift across the field of view may bias the likelihood.
        
    \subsection{Coherent GW search limits}
    \label{sec:results:injection-study}

        Here, we present our sensitivity to coherent GWs for the \kepler{} and \romanname{}  photometric surveys.
        We do this by reporting the minimum coherent GW strain $\strain$ for which the Bayes factor in comparison to the ``no coherent GW" case $\bayesf{}$ is above or equal to 10, for a collection of mock coherent GW sources.
        While the $\bayesf = 10$ threshold is entirely subjective, it is typically qualitatively taken to mean ``strong evidence" in favor of a given model \cite{kass+95_bayesfactortable}.

        Due to the small field of view of both surveys, we find that besides $\strain$, the primary factor in the detectability of a coherent GW source is the angular separation between the source and the field of view.
        Due to the quadrupole symmetry of the deflection response, for a given ring of constant angular deflection from a coherent GW source, the magnitude of observed deflections averaged across a field of view across one GW cycle is constant.
        The source-field of view angular separation (hereafter ``source-telescope separation", or just ``separation") produces a large effect on the strength of observed deflections.
        For the \keplerfov{} \kepler{} field of view, the average GW deflection length over one GW cycle for a $90 \degree$ source-telescope separation is $\sim 13$ times greater than a $0 \degree$ separation, marginalized over all parameters other than $\strain$ and source position.
        Due to the much smaller field of view (\romanfov{}) of \romanname{}, this effect is correspondingly greater, at $\sim 300 \times$.
        In order to account for this dependence, we generate a set of GW sources which is evenly distributed across the sky.

        We generate mock data for a grid of 20 frequencies across the frequency prior.
        For each frequency, we place GW sources at 16 evenly spaced positions on the celestial sphere, placed using a Fibonacci spiral centered on the field of view.
        For all position/frequency combinations, we generate a series of sources with 10 different trial strains.
        These trial strains are evenly spaced in $\logstrain$ along 1 order of magnitude, and centered at our best guess for the strain that produces a Bayes factor of 10.
        In summary, for each survey, our full mock run consists of 20 frequencies $\times$ 16 source positions $\times$ 10 strains.

        We generate our mock data using the \code{estoiles} code \cite{wang+21_romangw_coherent, wang+22_romangw_background}, and using the Newtonian (0PN) approximation.
        We add the astrometric white noise prescription detailed in Section \ref{sec:methods:data}.
        The 0PN approximation includes frequency evolution, so for a given strain, the exact combination of $\freq$, $\chirpmass$, and $\lumdist$ matters.
        For each generated source, $\freq$ is already fixed. We then select the minimum $\chirpmass$ for that $\freq$ such that the frequency does not evolve too rapidly.
        We define the speed of frequency evolution in terms of the GW phase difference between 0PN and our forward model, which assumes a constant frequency $\freq{}$ across the observation window.
        We pick the maximum $\chirpmass$ where this phase error remains below $\frac{\pi}{2}$ across the total survey observing time.
        $\lumdist$ is finally fixed to produce the desired strain.

        The remaining coherent GW parameters that we must decide on for our mock sources are the orbital inclination $\cosinc$, the orbital polarization angle $\polarizationangle$, and the initial GW phase $\initphase$.
        As we expect to view at least one GW cycle for our entire frequency range, our choices of $\polarizationangle$ and $\initphase$ should have only minor effect on the average deflection in our mock data, so we fix both to 0 with respect to the ICRS coordinate system.
        For computational reasons, we run the full frequency $\times$ position $\times$ strain grid above for only face-on inclinations ($\cosinc = 1$).
        As the orbital inclination approaches $\cosinc = 0$ (edge-on), average deflections decrease.
        We expect the threshold strain to increase as a result.
        We perform a subset of runs for \kepler{} to verify this, at $\freq{} = 10^{-7} \unit{Hz}$.
        For an inclination of $45\degree$ ($\cosinc{} = 0.5$), the $\bayesf{} = 10$ threshold strain increases by $\Delta \logstrain{} \approx 0.2$, and a similar increase occurs going from $45 \degree$ inclination to edge-on.

        Also for computational reasons, we do not generate data for the full $N_{\text{all}} = \keplernstars{}$ and $\romannstars{}$ stars for \kepler{} and \romanname{}  respectively.
        We instead generate deflections for a subset of $N_{\text{sub}}= \subsetnnstars{}$ stars, and scale down the astrometric noise by a factor of $\sqrt{N_{\text{sub}} / N_{\text{all}}}$.
        For all runs, we adopt an approximate likelihood using the frequency interpolation scheme described in Section $\ref{sec:methods:inner-prod}$, with $F = 1000$ interpolation grid points evenly spaced in $\logfreq{}$ over our frequency prior.
        While this $F$ is below the values needed to accurately approximate the evidence that we find in Section \ref{sec:results:inner-prod}, we emphasize that those results are for the actual number of stars in the dataset $N_{\text{all}}$.
        As the evidence accuracy scales as $\sqrt{N}$, since we work with a noise-scaled subset of size $N_{\text{sub}}$, $F = 1000$ is more than enough to well-approximate the likelihood.

        For \romanname{}, we consider the ``full mean-subtraction" case, where the mean pointing is fully absorbed by the pointing control system of \romanname{} \cite{wang+21_romangw_coherent, wang+22_romangw_background, pardo_2023_GravitationalWaveDetectiona}.
        This is predicted to decrease the sensitivity by a factor of 100 compared to an ideal case in which no pointing corrections are ever required \cite{wang+21_romangw_coherent}.
        In order to simulate full mean-subtraction in the data, we artificially increase the astrometric noise for Roman by this factor of 100.

        \begin{figure*}
            \centering
            \includegraphics[width=0.5\textwidth]{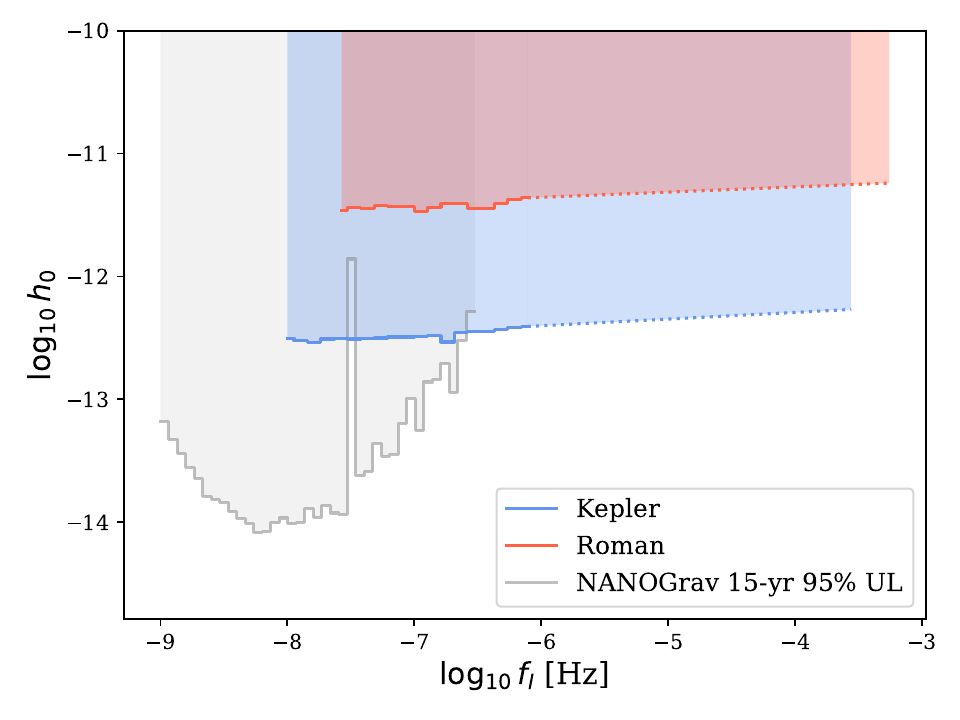}
            \caption{Position-averaged instantaneous strain sensitivity for \kepler{} and \romanname{}.
            These curves indicate the injected strain at which nested sampling yields a Bayes factor $\bayesf{} = 10$, averaged over a set of coherent GW source positions evenly distributed over the sky.
            We extrapolate our sensitivity results using a linear fit (in log-log space) from $10^{-6}$ $\unit{Hz}$ up to the Nyquist limit.
            We currently exclude this high-frequency region from our analysis due to the need to account for significant source frequency evolution over our observation window.
            We compare our results with 
            individual-source strain upper limits from the NANOGrav 15-year data release \cite{agazie_2023_NANOGrav15yearData_coherentsearch}.
            However, we caution that this curve comes from 95$\%$ strain upper limits on posteriors from the 15-year dataset, instead of the $\bayesf{} = 10$ threshold methodology we use.}
            \label{fig:coherent-strain-limits-sens-curve}
        \end{figure*}

        The results for these runs for \kepler{} and \romanname{} are shown in Figure \ref{fig:coherent-strain-limits-sens-curve}.
        The $\bayesf{} = 10$ curve reported for Kepler shows the true instantaneous source strain which results in a Bayes factor of 10 for each frequency bin.
        This is averaged over all GW source locations in each frequency bin.
        As the Fibonacci spiral used to generate the source locations divides the sky into approximately equal-area cells, we do no additional weighting for this average.

        \refresp{We expect our strain sensitivity to increase in proportion to the square root of the total number of observed data points ($N$ stars $\times$ $T$ observations per star), as the central limit theorem applies for the large number of observations we have for both \romanname{} and \kepler{}.
        In addition, we ordinarily would expect our sensitivity to increase with increasing frequency, from a similar argument to matched-filtering: we see more cycles of the GW (and are still well above the Nyquist sampling limit).
        However, interestingly we find that our strain sensitivity slightly decreases with increasing source frequency for both surveys.
        This decrease in sensitivity appears to stem from our adoption of a uniform log-frequency prior.
        For higher and higher injected GW frequencies, the likelihood spike in frequency space becomes narrower and narrower.
        Due to the nature of nested sampling, narrower spikes are easier to miss during the sampling process, leading to a higher $\bayesf{} = 10$ threshold strain on average.
        We have performed extensive investigations and confirmed that this effect persists even if no frequency evolution is allowed for the injected GWs, and for various combinations of nested sampling tuning parameters.
        While this sensitivity decrease vanishes if we choose to use a linearly-uniform prior for frequency, we believe choosing a linearly-uniform prior is not reflective of our expectations for a realistic population of coherent GW sources, as this prior heavily weights for GW frequencies close to the upper frequency cutoff.
        For a population of supermassive black hole binary GW sources, we instead might expect higher GW frequencies to be observed more rarely, as the chirp time decreases with higher frequencies.}
        
        \refresp{However, while we tentatively extrapolate our sensitivity to the higher-frequency region from $10^{-6}$ $\unit{Hz}$ to around $10^{-3.5}$ $\unit{Hz}$ with a power-law fit in Figure \ref{fig:coherent-strain-limits-sens-curve}, for a realistic search we must develop new methods for efficient coherent GW searches in this frequency band that take into account the significant levels of frequency evolution present.
        We expect the actual sensitivity from these new methods to differ from, and possibly improve on, the power-law extrapolation shown here.} %

        Averaging over all frequency bins (excluding the extrapolated high-frequency range), our mean instantaneous strain sensitivity for \kepler{} is $\strain{} \geq \kepleravgstrainlim$, and $\strain{} \geq \romanavgstrainlim$ for \romanname{}.
        For reference, we calculate a corresponding luminosity distance for a source with a plausible chirp mass and frequency.
        For such a source, with $\freq{} = 10^{-7}$ $\unit{Hz}$ and $\chirpmass{} = 10^9$ $\Msun{}$, the mean detectable $\lumdist{}$ for \kepler{} would be \kepleravglumdist{}, and \romanavglumdist{} for \romanname{}.
        We emphasize that these distances, being based on a strain sensitivity averaged over both source frequency and sky position, do not represent the maximum detectable luminosity distance for this source.
        Due to the strong dependence of GW deflection strength on the relative angular source-telescope separation, we expect to detect sources further than this mean distance given a favorable (i.e. $90\degree$ source-telescope separation).
        In addition, we reiterate that these limits are based on the assumption that a coherent GW is emitted from a slowly-inspiraling binary.
        A merger of a supermassive black hole binary would be detectable from much further away, given appropriate analysis.

        We compare our results to a sensitivity curve from the NANOGrav 15-year analysis.
        However, we caution that is not possible to make a direct comparison between our sensitivity curve, which is based on the Bayes factor, and this previous curve.
        For the NANOGrav sensitivity, a coherent GW Bayesian analysis is run on the 15-year dataset; finding no evidence for a coherent GW detection based on Bayes factors, the 95th percentile of the marginalized strain posteriors obtained from the analysis is taken as an upper limit on possible coherent GW strains.

        \refresp{Ref.~\cite{wang+21_romangw_coherent} also obtains a projected sensitivity for coherent GWs through a noise-scaling argument, where the limiting strain is taken as $h_{0,\text{lim}} = \frac{\sigma}{\sqrt{NT}}$, the average noise divided by the square root of the number of stars times the number of observations per star.
        While as previously mentioned we believe this sensitivity scaling relation is valid for \romanname{} and \kepler{} due to the central limit theorem, we do not directly include this limit for comparison in Figure \ref{fig:coherent-strain-limits-sens-curve}.
        This is because the limit given by Ref.~\cite{wang+21_romangw_coherent} must be adjusted to account for the average on-sky deflection per unit GW strain. 
        In addition, that limit, as published in Fig.~3 of Ref.~\cite{wang+21_romangw_coherent},  includes an additional constant factor used to compare to the characteristic strain sensitivities of other experiments. When these factors are corrected for, Ref.~\cite{wang+21_romangw_coherent}'s limit is comparable to ours.
        }

\begin{figure*}
            \centering
            \includegraphics[width=0.9\linewidth]{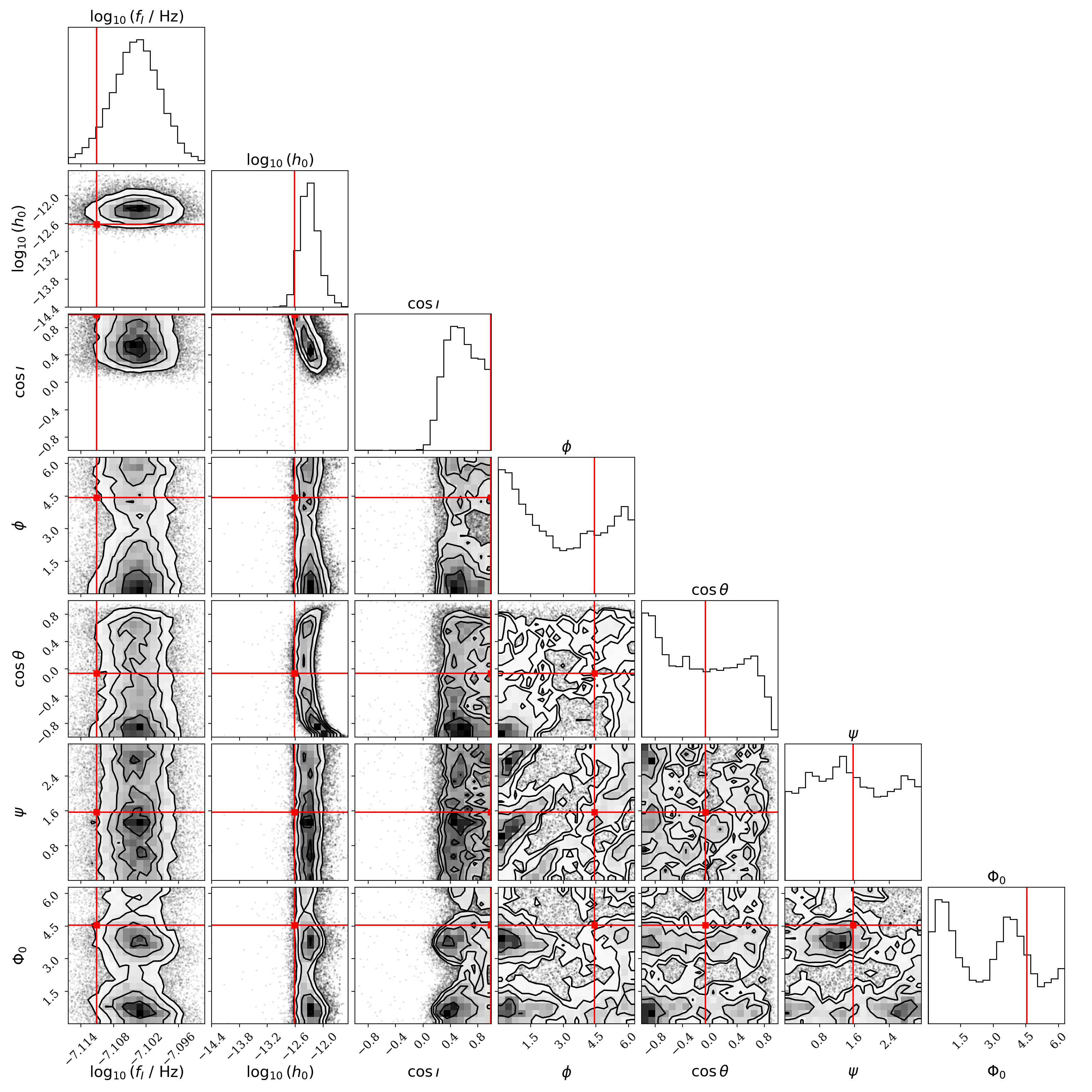}
            \caption{Corner plot of an individual nested sampling run for \kepler{}, at $\bayesf{} \sim 10$. This particular run is at $93\degree$ source-telescope separation, and $\freq{} = 10^{-7.11}$ $\unit{Hz}$. The red lines indicate the true GW parameters.
            The true values for $\psi$ and $\initphase$ differ from the values of zero quoted in Section \ref{sec:results:injection-study} due to the difference in coordinate systems for the GW source position; instead of being in ICRS, the coordinates $(\phi, \theta)$ are defined so that the telescope field of view center is located at $\cos \theta = 1$.
            The GW frequency is slightly overestimated compared to its true value due to the small amount of frequency evolution present in the mock data.
            Due to a strong degeneracy between $\strain{}$ and $\cosinc{}$, the marginalized posterior for the former overestimates the true value. However, this is largely due to the face-on inclination chosen for this source.}
            \label{fig:coherent-nontargeted-example}
        \end{figure*}

        Figure \ref{fig:coherent-strain-limits-sens-curve} abstracts away each individual Bayesian search.
        We therefore show a representative individual search with $\bayesf{} \sim 10$ for \kepler{} in Figure \ref{fig:coherent-nontargeted-example}, in order to discuss interesting features.
        This run is at a source-telescope separation of $93\degree$, and at a GW frequency of $\freq{} = 10^{-7.11}$ $\unit{Hz}$.
        Although we find that the posterior of $\logfreq{}$ is constrained to good precision, the small amount of frequency evolution in the mock data causes it to be consistently overestimated compared to the true value (in red).
        We observe a strong degeneracy between $\logstrain{}$ and $\cosinc{}$, though the recovered posterior contains the true $\logstrain{}$-$\cosinc{}$ combination within the 2$\sigma$ contour.
        Due to the small survey field of view, we achieve extremely poor source localization despite good recovery of $\logfreq$ and $\logstrain$.
        Interestingly, the recovered posterior for $\cos \theta$ is more biased towards the field's antipode (at $\cos \theta = -1$) rather than its center.
        We believe that this is due to the differing curl properties of deflections for sources at the field vs. antipodal sky hemisphere.
        Deflections for sources within the antipodal hemisphere visually appear more similar to deflections at $90\degree$ separation than sources within the field-centered hemisphere do.
        As discussed in Section \ref{sec:methods:bayesian-search}, the degeneracy between $\polarizationangle$ and $\initphase$ for a face-on inclination manifests as separate peaks in the posterior, due to the artificial prior limits imposed on these angular quantities.

    \subsection{Targeted coherent GW search limits}

        We explore our relative sensitivity to coherent GW sources where we have a known source position versus when the source position is left as a free parameter.
        This can be the case when conducting searches from candidates found in EM searches, but also forms part of a general search strategy.
        We find in the last section that the maximum luminosity distance for even sources on the high-mass, high-frequency end is relatively short (for \kepler{}, \kepleravglumdist{} for a source with $\freq{} = 10^{-7} \unit{Hz}$ and $\chirpmass{} = 10^9 \Msun$).
        Therefore, we can consider adopting a search strategy where the prior on source positions is not continuously uniform across the entire sky, but is discrete, on galactic positions taken from a suitable catalog.
        Nested sampling allows us to implement this straightforwardly, as a nested sampling run yields the Bayesian evidence.
        We can therefore conduct one nested sampling run per candidate galaxy, with the source on-sky location fixed to the candidate's, and then combine the obtained posteriors by reweighting using the evidence.
        The recombined posterior on the location will be a discrete probability distribution.

        To quantify the effectiveness of this strategy, we rerun our $\bayesf{} = 10$ threshold strain analysis for $\kepler{}$ at the frequency $\freq{} = 10^{-7} \unit{Hz}$, with the source position fixed to truth.
        The Bayes factor obtained from these runs is an upper bound on any Bayes factor from the discrete position prior search described above, as it is likely that the evidence from the run at the true position will be ``diluted" by runs at all other candidate positions when calculating the combined evidence.
        
\begin{figure*}
            \centering
            \includegraphics[width=0.9\linewidth]{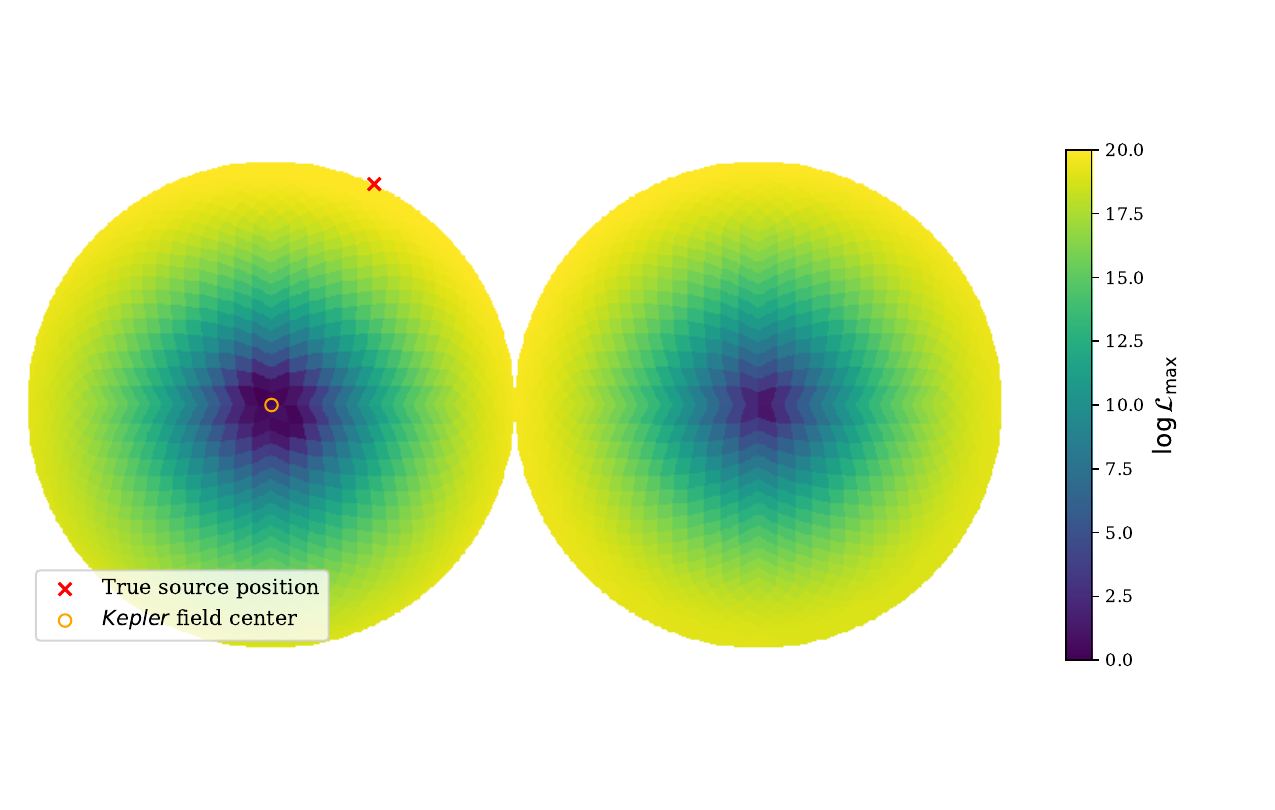}
            \caption{Polar-projection plot of maximum likelihood versus source position, for a \kepler{} $90\degree$-separation source at $\bayesf{} \sim 10$. The hemispheres are centered on the telescope field of view center and its antipode. For every (fixed) source position on this HEALPix grid, we maximize the likelihood with respect to all other parameters. Due to the small field of view of \kepler{}, we see that across most of the sky, the likelihood is nearly equal to the maximum likelihood at the true source position.}
            \label{fig:coherent-likelihood-position}
        \end{figure*}

        Unfortunately, we find that fixing the source position only provides a marginal improvement in sensitivity, with the threshold strain decreasing by $\Delta \logstrain{} \approx -0.2$.
        Counterintuitively, this is likely due to the poor source localization caused by our small fields of view.
        For large angular separations, which are the most commonly expected, the maximum likelihood remains similar across most of the sky, when marginalizing over the source position only.
        Importantly, this implies that the maximum likelihood across most of the sky is also similar to the maximum likelihood at the true source position.
        This is illustrated in Figure \ref{fig:coherent-likelihood-position}, for a source at $90 \degree$ separation.
        The Bayesian evidence is defined as the prior-weighted average of the likelihood across all parameter space; therefore, the source position-marginalized evidence for a likelihood which is approximately equal to the likelihood at the true source position, will be similar to the source position-fixed evidence.
        Despite these mixed results, given the ease of assembling a complete galaxy catalog within our maximum detectable $\lumdist{}$, performing a position-targeted search may still be worth doing on actual data.

        We can also consider a targeted search in source frequency as well as position.
        For PTAs, Ref.~\cite{arzoumanian+20} study the effect of frequency targeting for the SMBHB host galaxy candidate 3C 66B.
        They find that shrinking the width of the frequency prior by $0.75$ orders of magnitude using limits from EM studies results in an order of magnitude decrease in the upper chirp mass limit.
        We are confident that shrinking our frequency prior width will provide similar improvements, as we observe no covariance between the recovered frequency and source localization in Figure \ref{fig:coherent-nontargeted-example}.
        However, explicitly confirming this result for coherent GW searches using astrometric deflections is beyond the scope of this work.

\section{Conclusion}

    In this work, we explore the prospects of detecting coherent gravitational wave sources from the \kepler{} and \romanname{}  photometric surveys, using relative stellar astrometry.
    Unlike any other GW detection method, for astrometry we have many ($10^5-10^8$) baselines, with an extremely low SNR per baseline.
    This presents a unique computational challenge for conventional Bayesian signal searches due to the large (terabyte-scale) raw dataset size.
    In order to overcome this challenge, we extend a technique developed for pulsar timing arrays which speeds up Bayesian likelihood evaluations by pre-computing inner products between the data and candidate signal templates.
    We find that by operating on these inner products, we can reduce the dataset size by a factor of $\mathcal{O}(10)-\mathcal{O}(100)$ with a $<1\%$ reduction in accuracy.
    
    Using a comprehensive GW signal injection strategy, we find that \kepler{} offers the best prospects for coherent GW detection due to its surveying strategy, with an instantaneous strain limit of $\strain{} \geq \kepleravgstrainlim{}$.
    In comparison, although \romanname{}  has a comparable number of observations and a factor of $\sim 10^3$ more stars, the absorption of the mean GW deflection through systematics means that its limiting strain is a factor of $\sim 10$ smaller, at $\strain{} \geq \romanavgstrainlim{}$.
    Due to the small fields of view of these surveys, we find that though frequency and strain can be well-constrained, sources are generally poorly localized on the sky.
    Given the relatively small ($\lesssim 10$ Mpc) distances accessible to us, we assess whether sensitivity can be improved by targeting individual candidate galaxy positions.
    Unfortunately, due to poor source localization, we find that the improvement from targeting individual galaxy positions is relatively minor.
    However, we expect to achieve more substantial improvements in sensitivity given frequency prior constraints from electromagnetic studies.

    The primary limitation of our analysis in our work is the assumption that the frequency of the coherent GWs we consider does not appreciably evolve within the total observing time of our surveys, which is necessary for the inner product precomputation technique.
    We artificially impose this assumption when forecasting results for the high end ($\sim 10^{-6}$ $\unit{Hz}$) of the frequency range we consider in this work.
    However, even for microhertz frequencies, this requires imposing an unrealistically low chirp mass on our sources, with a correspondingly unrealistically low ($<0.1$ Mpc) luminosity distance for the strains to which we are sensitive.
    In practice, the high cadence of our observations means that we expect to also be sensitive to frequencies in the range $10^{-6}$ $\unit{Hz}-10^{-4}$ $\unit{Hz}$.
    Any sources we encounter within this frequency range are almost certainly rapidly upchirping on timescales of months to days.
    We therefore hope to extend our dataset compression scheme to encompass rapidly frequency-evolving waveforms in future work.
    
    \refresp{In particular, a significant amount of work has been devoted to reducing the computational cost of searching for GW mergers in the ground-based LIGO/KAGRA/VIRGO frequency band, given the stringent latency requirements for followup electromagnetic observations.
    Methods such as reduced-order quadratures \citep{antil_2012_roq} and heterodyning \citep{leslie_2021_ModebymodeRelativeBinning, cornish_2021_HeterodynedLikelihoodRapid} can shrink datasets by a factor of a hundred with essentially zero loss in accuracy.
    However, these techniques assume that the coherent GW inspiral-merger-ringdown waveform, while complicated, is observed over a short time over low noise conditions compared to astrometry.
    We expect to encounter novel challenges when adapting these LIGO-band techniques to the astrometric regime, where instead a much simpler inspiral waveform is observed over many cycles, with a high noise level, and in many stellar baselines.}

    Although GW detection via astrometry presents a computational challenge, it offers a promising method for bridging the frequency gap of $10^{-7}-10^{-4}$ $\unit{Hz}$ between pulsar timing arrays and direct detection methods.
    Using serendipitous data from the past \kepler{} and near-future \romanname{} missions, we hope to directly probe this frequency range, which may contain GW sources of both astrophysical and cosmological interest.

\begin{acknowledgments}
    The authors would like to thank Bence B\'{e}csy and Michele Vallisneri for valuable discussions related to this work at the IPTA 2024 meeting. 
    \refresp{This work makes use of the gaia-kepler.fun crossmatch database created by Megan Bedell.}
    The authors acknowledge funding support from the NASA ROSES ADAP grant 80NSSC23K0629 and the NASA ROSES Roman grant 22-ROMAN22-0040. The authors also acknowledge the Center for Advanced Research Computing (CARC) at the University of Southern California for providing computing resources that have contributed to the research results reported within this publication. Part of this work was done at Jet Propulsion Laboratory, California Institute of Technology, under a contract with the National Aeronautics and Space Administration (80NM0018D0004). This work was supported by the Carnegie Institution for Science’s Carnegie Fellowship (LGB).
\end{acknowledgments}

\appendix

\section{Nested sampling tuning parameters}
\label{appendix:nested-sampling}

    For the nested sampling runs we perform in Section \ref{sec:results:injection-study}, we use version 2.5.1 of JAXNS \cite{albert20_jaxns1, albert23_jaxns2}.
    We use $10^4$ live points, and terminate if the estimated remaining evidence $\mathrm{d}\ln{(\evidence{})}$ is below $\ln{(1 + 10^{-3})}$.
    We enable the $\code{difficult\_model}$ and $\code{parameter\_estimation}$ flags, which provide more robust default nested sampling settings.

\bibliography{main}%

\providecommand{\noopsort}[1]{}\providecommand{\singleletter}[1]{#1}%
\begin{thebibliography}{42}%
\makeatletter
\providecommand \@ifxundefined [1]{%
 \@ifx{#1\undefined}
}%
\providecommand \@ifnum [1]{%
 \ifnum #1\expandafter \@firstoftwo
 \else \expandafter \@secondoftwo
 \fi
}%
\providecommand \@ifx [1]{%
 \ifx #1\expandafter \@firstoftwo
 \else \expandafter \@secondoftwo
 \fi
}%
\providecommand \natexlab [1]{#1}%
\providecommand \enquote  [1]{``#1''}%
\providecommand \bibnamefont  [1]{#1}%
\providecommand \bibfnamefont [1]{#1}%
\providecommand \citenamefont [1]{#1}%
\providecommand \href@noop [0]{\@secondoftwo}%
\providecommand \href [0]{\begingroup \@sanitize@url \@href}%
\providecommand \@href[1]{\@@startlink{#1}\@@href}%
\providecommand \@@href[1]{\endgroup#1\@@endlink}%
\providecommand \@sanitize@url [0]{\catcode `\\12\catcode `\$12\catcode `\&12\catcode `\#12\catcode `\^12\catcode `\_12\catcode `\%12\relax}%
\providecommand \@@startlink[1]{}%
\providecommand \@@endlink[0]{}%
\providecommand \url  [0]{\begingroup\@sanitize@url \@url }%
\providecommand \@url [1]{\endgroup\@href {#1}{\urlprefix }}%
\providecommand \urlprefix  [0]{URL }%
\providecommand \Eprint [0]{\href }%
\providecommand \doibase [0]{https://doi.org/}%
\providecommand \selectlanguage [0]{\@gobble}%
\providecommand \bibinfo  [0]{\@secondoftwo}%
\providecommand \bibfield  [0]{\@secondoftwo}%
\providecommand \translation [1]{[#1]}%
\providecommand \BibitemOpen [0]{}%
\providecommand \bibitemStop [0]{}%
\providecommand \bibitemNoStop [0]{.\EOS\space}%
\providecommand \EOS [0]{\spacefactor3000\relax}%
\providecommand \BibitemShut  [1]{\csname bibitem#1\endcsname}%
\let\auto@bib@innerbib\@empty
\bibitem [{\citenamefont {{Abbott}}\ \emph {et~al.}(2016)\citenamefont {{Abbott}}, \citenamefont {{Abbott}}, \citenamefont {{Abbott}}, \citenamefont {{Abernathy}}, \citenamefont {{Acernese}}, \citenamefont {{Ackley}}, \citenamefont {{Adams}}, \citenamefont {{Adams}}, \citenamefont {{Addesso}}, \citenamefont {{Adhikari}},\ and\ \citenamefont {et~al.}}]{GW150914}%
  \BibitemOpen
  \bibfield  {author} {\bibinfo {author} {\bibfnamefont {B.~P.}\ \bibnamefont {{Abbott}}}, \bibinfo {author} {\bibfnamefont {R.}~\bibnamefont {{Abbott}}}, \bibinfo {author} {\bibfnamefont {T.~D.}\ \bibnamefont {{Abbott}}}, \emph {et~al.},\ }\href {https://doi.org/10.1103/PhysRevD.93.122003} {\bibfield  {journal} {\bibinfo  {journal} {\prd}\ }\textbf {\bibinfo {volume} {93}},\ \bibinfo {eid} {122003} (\bibinfo {year} {2016})},\ \Eprint {https://arxiv.org/abs/1602.03839} {arXiv:1602.03839 [gr-qc]} \BibitemShut {NoStop}%
\bibitem [{\citenamefont {Agazie}\ \emph {et~al.}(2023{\natexlab{a}})\citenamefont {Agazie}, \citenamefont {Anumarlapudi}, \citenamefont {Archibald}, \citenamefont {Arzoumanian}, \citenamefont {Baker}, \citenamefont {Bécsy}, \citenamefont {Blecha}, \citenamefont {Brazier}, \citenamefont {Brook}, \citenamefont {Burke-Spolaor}, \citenamefont {Burnette}, \citenamefont {Case}, \citenamefont {Charisi}, \citenamefont {Chatterjee}, \citenamefont {Chatziioannou}, \citenamefont {Cheeseboro}, \citenamefont {Chen}, \citenamefont {Cohen}, \citenamefont {Cordes}, \citenamefont {Cornish}, \citenamefont {Crawford}, \citenamefont {Cromartie}, \citenamefont {Crowter}, \citenamefont {Cutler}, \citenamefont {DeCesar}, \citenamefont {DeGan}, \citenamefont {Demorest}, \citenamefont {Deng}, \citenamefont {Dolch}, \citenamefont {Drachler}, \citenamefont {Ellis}, \citenamefont {Ferrara}, \citenamefont {Fiore}, \citenamefont {Fonseca}, \citenamefont {Freedman}, \citenamefont {Garver-Daniels}, \citenamefont {Gentile}, \citenamefont
  {Gersbach}, \citenamefont {Glaser}, \citenamefont {Good}, \citenamefont {Gültekin}, \citenamefont {Hazboun}, \citenamefont {Hourihane}, \citenamefont {Islo}, \citenamefont {Jennings}, \citenamefont {Johnson}, \citenamefont {Jones}, \citenamefont {Kaiser}, \citenamefont {Kaplan}, \citenamefont {Kelley}, \citenamefont {Kerr}, \citenamefont {Key}, \citenamefont {Klein}, \citenamefont {Laal}, \citenamefont {Lam}, \citenamefont {Lamb}, \citenamefont {Lazio}, \citenamefont {Lewandowska}, \citenamefont {Littenberg}, \citenamefont {Liu}, \citenamefont {Lommen}, \citenamefont {Lorimer}, \citenamefont {Luo}, \citenamefont {Lynch}, \citenamefont {Ma}, \citenamefont {Madison}, \citenamefont {Mattson}, \citenamefont {McEwen}, \citenamefont {McKee}, \citenamefont {McLaughlin}, \citenamefont {McMann}, \citenamefont {Meyers}, \citenamefont {Meyers}, \citenamefont {Mingarelli}, \citenamefont {Mitridate}, \citenamefont {Natarajan}, \citenamefont {Ng}, \citenamefont {Nice}, \citenamefont {Ocker}, \citenamefont {Olum},
  \citenamefont {Pennucci}, \citenamefont {Perera}, \citenamefont {Petrov}, \citenamefont {Pol}, \citenamefont {Radovan}, \citenamefont {Ransom}, \citenamefont {Ray}, \citenamefont {Romano}, \citenamefont {Sardesai}, \citenamefont {Schmiedekamp}, \citenamefont {Schmiedekamp}, \citenamefont {Schmitz}, \citenamefont {Schult}, \citenamefont {Shapiro-Albert}, \citenamefont {Siemens}, \citenamefont {Simon}, \citenamefont {Siwek}, \citenamefont {Stairs}, \citenamefont {Stinebring}, \citenamefont {Stovall}, \citenamefont {Sun}, \citenamefont {Susobhanan}, \citenamefont {Swiggum}, \citenamefont {Taylor}, \citenamefont {Taylor}, \citenamefont {Turner}, \citenamefont {Unal}, \citenamefont {Vallisneri}, \citenamefont {van Haasteren}, \citenamefont {Vigeland}, \citenamefont {Wahl}, \citenamefont {Wang}, \citenamefont {Witt}, \citenamefont {Young},\ and\ \citenamefont {Collaboration}}]{agazie+23_nanograv_15yr_gwb}%
  \BibitemOpen
  \bibfield  {author} {\bibinfo {author} {\bibfnamefont {G.}~\bibnamefont {Agazie}}, \bibinfo {author} {\bibfnamefont {A.}~\bibnamefont {Anumarlapudi}}, \bibinfo {author} {\bibfnamefont {A.~M.}\ \bibnamefont {Archibald}}, \emph {et~al.},\ }\href {https://doi.org/10.3847/2041-8213/acdac6} {\bibfield  {journal} {\bibinfo  {journal} {The Astrophysical Journal Letters}\ }\textbf {\bibinfo {volume} {951}},\ \bibinfo {pages} {L8} (\bibinfo {year} {2023}{\natexlab{a}})}\BibitemShut {NoStop}%
\bibitem [{\citenamefont {{EPTA Collaboration and InPTA Collaboration:}}\ \emph {et~al.}(2023)\citenamefont {{EPTA Collaboration and InPTA Collaboration:}}, \citenamefont {Antoniadis}, \citenamefont {Arumugam}, \citenamefont {Arumugam}, \citenamefont {Babak}, \citenamefont {Bagchi}, \citenamefont {Bak~Nielsen}, \citenamefont {Bassa}, \citenamefont {Bathula}, \citenamefont {Berthereau}, \citenamefont {Bonetti}, \citenamefont {Bortolas}, \citenamefont {Brook}, \citenamefont {Burgay}, \citenamefont {Caballero}, \citenamefont {Chalumeau}, \citenamefont {Champion}, \citenamefont {Chanlaridis}, \citenamefont {Chen}, \citenamefont {Cognard}, \citenamefont {Dandapat}, \citenamefont {Deb}, \citenamefont {Desai}, \citenamefont {Desvignes}, \citenamefont {Dhanda-Batra}, \citenamefont {Dwivedi}, \citenamefont {Falxa}, \citenamefont {Ferdman}, \citenamefont {Franchini}, \citenamefont {Gair}, \citenamefont {Goncharov}, \citenamefont {Gopakumar}, \citenamefont {Graikou}, \citenamefont {Grießmeier}, \citenamefont
  {Guillemot}, \citenamefont {Guo}, \citenamefont {Gupta}, \citenamefont {Hisano}, \citenamefont {Hu}, \citenamefont {Iraci}, \citenamefont {Izquierdo-Villalba}, \citenamefont {Jang}, \citenamefont {Jawor}, \citenamefont {Janssen}, \citenamefont {Jessner}, \citenamefont {Joshi}, \citenamefont {Kareem}, \citenamefont {Karuppusamy}, \citenamefont {Keane}, \citenamefont {Keith}, \citenamefont {Kharbanda}, \citenamefont {Kikunaga}, \citenamefont {Kolhe}, \citenamefont {Kramer}, \citenamefont {Krishnakumar}, \citenamefont {Lackeos}, \citenamefont {Lee}, \citenamefont {Liu}, \citenamefont {Liu}, \citenamefont {Lyne}, \citenamefont {McKee}, \citenamefont {Maan}, \citenamefont {Main}, \citenamefont {Mickaliger}, \citenamefont {Niţu}, \citenamefont {Nobleson}, \citenamefont {Paladi}, \citenamefont {Parthasarathy}, \citenamefont {Perera}, \citenamefont {Perrodin}, \citenamefont {Petiteau}, \citenamefont {Porayko}, \citenamefont {Possenti}, \citenamefont {Prabu}, \citenamefont {Quelquejay~Leclere}, \citenamefont
  {Rana}, \citenamefont {Samajdar}, \citenamefont {Sanidas}, \citenamefont {Sesana}, \citenamefont {Shaifullah}, \citenamefont {Singha}, \citenamefont {Speri}, \citenamefont {Spiewak}, \citenamefont {Srivastava}, \citenamefont {Stappers}, \citenamefont {Surnis}, \citenamefont {Susarla}, \citenamefont {Susobhanan}, \citenamefont {Takahashi}, \citenamefont {Tarafdar}, \citenamefont {Theureau}, \citenamefont {Tiburzi}, \citenamefont {Van Der~Wateren}, \citenamefont {Vecchio}, \citenamefont {Venkatraman~Krishnan}, \citenamefont {Verbiest}, \citenamefont {Wang}, \citenamefont {Wang},\ and\ \citenamefont {Wu}}]{eptacollaborationandinptacollaboration:_2023_SecondDataRelease}%
  \BibitemOpen
  \bibfield  {author} {\bibinfo {author} {\bibnamefont {{EPTA Collaboration and InPTA Collaboration:}}}, \bibinfo {author} {\bibfnamefont {J.}~\bibnamefont {Antoniadis}}, \bibinfo {author} {\bibfnamefont {P.}~\bibnamefont {Arumugam}}, \emph {et~al.},\ }\href {https://doi.org/10.1051/0004-6361/202346844} {\bibfield  {journal} {\bibinfo  {journal} {Astronomy \& Astrophysics}\ }\textbf {\bibinfo {volume} {678}},\ \bibinfo {pages} {A50} (\bibinfo {year} {2023})}\BibitemShut {NoStop}%
\bibitem [{\citenamefont {Xu}\ \emph {et~al.}(2023)\citenamefont {Xu}, \citenamefont {Chen}, \citenamefont {Guo}, \citenamefont {Jiang}, \citenamefont {Wang}, \citenamefont {Xu}, \citenamefont {Xue}, \citenamefont {Nicolas~Caballero}, \citenamefont {Yuan}, \citenamefont {Xu}, \citenamefont {Wang}, \citenamefont {Hao}, \citenamefont {Luo}, \citenamefont {Lee}, \citenamefont {Han}, \citenamefont {Jiang}, \citenamefont {Shen}, \citenamefont {Wang}, \citenamefont {Wang}, \citenamefont {Xu}, \citenamefont {Wu}, \citenamefont {Manchester}, \citenamefont {Qian}, \citenamefont {Guan}, \citenamefont {Huang}, \citenamefont {Sun},\ and\ \citenamefont {Zhu}}]{xu_2023_SearchingNanoHertzStochastic}%
  \BibitemOpen
  \bibfield  {author} {\bibinfo {author} {\bibfnamefont {H.}~\bibnamefont {Xu}}, \bibinfo {author} {\bibfnamefont {S.}~\bibnamefont {Chen}}, \bibinfo {author} {\bibfnamefont {Y.}~\bibnamefont {Guo}}, \emph {et~al.},\ }\href {https://doi.org/10.1088/1674-4527/acdfa5} {\bibfield  {journal} {\bibinfo  {journal} {Research in Astronomy and Astrophysics}\ }\textbf {\bibinfo {volume} {23}},\ \bibinfo {pages} {075024} (\bibinfo {year} {2023})}\BibitemShut {NoStop}%
\bibitem [{\citenamefont {Reardon}\ \emph {et~al.}(2023)\citenamefont {Reardon}, \citenamefont {Zic}, \citenamefont {Shannon}, \citenamefont {Hobbs}, \citenamefont {Bailes}, \citenamefont {Di~Marco}, \citenamefont {Kapur}, \citenamefont {Rogers}, \citenamefont {Thrane}, \citenamefont {Askew}, \citenamefont {Bhat}, \citenamefont {Cameron}, \citenamefont {Curyło}, \citenamefont {Coles}, \citenamefont {Dai}, \citenamefont {Goncharov}, \citenamefont {Kerr}, \citenamefont {Kulkarni}, \citenamefont {Levin}, \citenamefont {Lower}, \citenamefont {Manchester}, \citenamefont {Mandow}, \citenamefont {Miles}, \citenamefont {Nathan}, \citenamefont {Osłowski}, \citenamefont {Russell}, \citenamefont {Spiewak}, \citenamefont {Zhang},\ and\ \citenamefont {Zhu}}]{reardon_2023_SearchIsotropicGravitationalwave}%
  \BibitemOpen
  \bibfield  {author} {\bibinfo {author} {\bibfnamefont {D.~J.}\ \bibnamefont {Reardon}}, \bibinfo {author} {\bibfnamefont {A.}~\bibnamefont {Zic}}, \bibinfo {author} {\bibfnamefont {R.~M.}\ \bibnamefont {Shannon}}, \emph {et~al.},\ }\href {https://doi.org/10.3847/2041-8213/acdd02} {\bibfield  {journal} {\bibinfo  {journal} {The Astrophysical Journal Letters}\ }\textbf {\bibinfo {volume} {951}},\ \bibinfo {pages} {L6} (\bibinfo {year} {2023})}\BibitemShut {NoStop}%
\bibitem [{\citenamefont {{Afzal}}\ \emph {et~al.}(2023)\citenamefont {{Afzal}}, \citenamefont {{Agazie}}, \citenamefont {{Anumarlapudi}}, \citenamefont {{Archibald}}, \citenamefont {{Arzoumanian}}, \citenamefont {{Baker}}, \citenamefont {{B{\'e}csy}}, \citenamefont {{Blanco-Pillado}}, \citenamefont {{Blecha}}, \citenamefont {{Boddy}}, \citenamefont {{Brazier}}, \citenamefont {{Brook}}, \citenamefont {{Burke-Spolaor}}, \citenamefont {{Burnette}}, \citenamefont {{Case}}, \citenamefont {{Charisi}}, \citenamefont {{Chatterjee}}, \citenamefont {{Chatziioannou}}, \citenamefont {{Cheeseboro}}, \citenamefont {{Chen}}, \citenamefont {{Cohen}}, \citenamefont {{Cordes}}, \citenamefont {{Cornish}}, \citenamefont {{Crawford}}, \citenamefont {{Cromartie}}, \citenamefont {{Crowter}}, \citenamefont {{Cutler}}, \citenamefont {{Decesar}}, \citenamefont {{Degan}}, \citenamefont {{Demorest}}, \citenamefont {{Deng}}, \citenamefont {{Dolch}}, \citenamefont {{Drachler}}, \citenamefont {{von Eckardstein}}, \citenamefont
  {{Ferrara}}, \citenamefont {{Fiore}}, \citenamefont {{Fonseca}}, \citenamefont {{Freedman}}, \citenamefont {{Garver-Daniels}}, \citenamefont {{Gentile}}, \citenamefont {{Gersbach}}, \citenamefont {{Glaser}}, \citenamefont {{Good}}, \citenamefont {{Guertin}}, \citenamefont {{G{\"u}ltekin}}, \citenamefont {{Hazboun}}, \citenamefont {{Hourihane}}, \citenamefont {{Islo}}, \citenamefont {{Jennings}}, \citenamefont {{Johnson}}, \citenamefont {{Jones}}, \citenamefont {{Kaiser}}, \citenamefont {{Kaplan}}, \citenamefont {{Kelley}}, \citenamefont {{Kerr}}, \citenamefont {{Key}}, \citenamefont {{Laal}}, \citenamefont {{Lam}}, \citenamefont {{Lamb}}, \citenamefont {{Lazio}}, \citenamefont {{Lee}}, \citenamefont {{Lewandowska}}, \citenamefont {{Lino Dos Santos}}, \citenamefont {{Littenberg}}, \citenamefont {{Liu}}, \citenamefont {{Lorimer}}, \citenamefont {{Luo}}, \citenamefont {{Lynch}}, \citenamefont {{Ma}}, \citenamefont {{Madison}}, \citenamefont {{McEwen}}, \citenamefont {{McKee}}, \citenamefont {{McLaughlin}},
  \citenamefont {{McMann}}, \citenamefont {{Meyers}}, \citenamefont {{Meyers}}, \citenamefont {{Mingarelli}}, \citenamefont {{Mitridate}}, \citenamefont {{Nay}}, \citenamefont {{Natarajan}}, \citenamefont {{Ng}}, \citenamefont {{Nice}}, \citenamefont {{Ocker}}, \citenamefont {{Olum}}, \citenamefont {{Pennucci}}, \citenamefont {{Perera}}, \citenamefont {{Petrov}}, \citenamefont {{Pol}}, \citenamefont {{Radovan}}, \citenamefont {{Ransom}}, \citenamefont {{Ray}}, \citenamefont {{Romano}}, \citenamefont {{Sardesai}}, \citenamefont {{Schmiedekamp}}, \citenamefont {{Schmiedekamp}}, \citenamefont {{Schmitz}}, \citenamefont {{Schr{\"o}der}}, \citenamefont {{Schult}}, \citenamefont {{Shapiro-Albert}}, \citenamefont {{Siemens}}, \citenamefont {{Simon}}, \citenamefont {{Siwek}}, \citenamefont {{Stairs}}, \citenamefont {{Stinebring}}, \citenamefont {{Stovall}}, \citenamefont {{Stratmann}}, \citenamefont {{Sun}}, \citenamefont {{Susobhanan}}, \citenamefont {{Swiggum}}, \citenamefont {{Taylor}}, \citenamefont {{Taylor}},
  \citenamefont {{Trickle}}, \citenamefont {{Turner}}, \citenamefont {{Unal}}, \citenamefont {{Vallisneri}}, \citenamefont {{Verma}}, \citenamefont {{Vigeland}}, \citenamefont {{Wahl}}, \citenamefont {{Wang}}, \citenamefont {{Witt}}, \citenamefont {{Wright}}, \citenamefont {{Young}}, \citenamefont {{Zurek}},\ and\ \citenamefont {{Nanograv Collaboration}}}]{afzal+2023_nanograv_15yr_bsm}%
  \BibitemOpen
  \bibfield  {author} {\bibinfo {author} {\bibfnamefont {A.}~\bibnamefont {{Afzal}}}, \bibinfo {author} {\bibfnamefont {G.}~\bibnamefont {{Agazie}}}, \bibinfo {author} {\bibfnamefont {A.}~\bibnamefont {{Anumarlapudi}}}, \emph {et~al.},\ }\href {https://doi.org/10.3847/2041-8213/acdc91} {\bibfield  {journal} {\bibinfo  {journal} {\apjl}\ }\textbf {\bibinfo {volume} {951}},\ \bibinfo {eid} {L11} (\bibinfo {year} {2023})},\ \Eprint {https://arxiv.org/abs/2306.16219} {arXiv:2306.16219 [astro-ph.HE]} \BibitemShut {NoStop}%
\bibitem [{\citenamefont {Agazie}\ \emph {et~al.}(2024)\citenamefont {Agazie}, \citenamefont {Baker}, \citenamefont {Bécsy}, \citenamefont {Blecha}, \citenamefont {Brazier}, \citenamefont {Brook}, \citenamefont {Brown}, \citenamefont {Burke-Spolaor}, \citenamefont {Casey-Clyde}, \citenamefont {Charisi}, \citenamefont {Chatterjee}, \citenamefont {Cohen}, \citenamefont {Cordes}, \citenamefont {Cornish}, \citenamefont {Crawford}, \citenamefont {Cromartie}, \citenamefont {DeCesar}, \citenamefont {Demorest}, \citenamefont {Deng}, \citenamefont {Dolch}, \citenamefont {Ferrara}, \citenamefont {Fiore}, \citenamefont {Fonseca}, \citenamefont {Freedman}, \citenamefont {Garver-Daniels}, \citenamefont {Glaser}, \citenamefont {Good}, \citenamefont {Gültekin}, \citenamefont {Hazboun}, \citenamefont {Jennings}, \citenamefont {Johnson}, \citenamefont {Jones}, \citenamefont {Kaiser}, \citenamefont {Kaplan}, \citenamefont {Kelley}, \citenamefont {Key}, \citenamefont {Laal}, \citenamefont {Lam}, \citenamefont {Lamb},
  \citenamefont {Larsen}, \citenamefont {Lazio}, \citenamefont {Lewandowska}, \citenamefont {Liu}, \citenamefont {Luo}, \citenamefont {Lynch}, \citenamefont {Ma}, \citenamefont {Madison}, \citenamefont {McEwen}, \citenamefont {McKee}, \citenamefont {McLaughlin}, \citenamefont {Meyers}, \citenamefont {Mingarelli}, \citenamefont {Mitridate}, \citenamefont {Natarajan}, \citenamefont {Nice}, \citenamefont {Ocker}, \citenamefont {Olum}, \citenamefont {Pennucci}, \citenamefont {Pol}, \citenamefont {Radovan}, \citenamefont {Ransom}, \citenamefont {Ray}, \citenamefont {Romano}, \citenamefont {Runnoe}, \citenamefont {Sardesai}, \citenamefont {Schmitz}, \citenamefont {Siemens}, \citenamefont {Simon}, \citenamefont {Siwek}, \citenamefont {Fiscella}, \citenamefont {Stairs}, \citenamefont {Stinebring}, \citenamefont {Susobhanan}, \citenamefont {Swiggum}, \citenamefont {Taylor}, \citenamefont {Turner}, \citenamefont {Unal}, \citenamefont {Vallisneri}, \citenamefont {Vigeland}, \citenamefont {Wahl}, \citenamefont {Willson},
  \citenamefont {Witt},\ and\ \citenamefont {Young}}]{agazie_2024_NANOGrav15Yr_discreteGWB}%
  \BibitemOpen
  \bibfield  {author} {\bibinfo {author} {\bibfnamefont {G.}~\bibnamefont {Agazie}}, \bibinfo {author} {\bibfnamefont {P.~T.}\ \bibnamefont {Baker}}, \bibinfo {author} {\bibfnamefont {B.}~\bibnamefont {Bécsy}}, \emph {et~al.},\ }\href {http://arxiv.org/abs/2404.07020} {\bibinfo {title} {The {NANOGrav} 15 yr {Data} {Set}: {Looking} for {Signs} of {Discreteness} in the {Gravitational}-wave {Background}}} (\bibinfo {year} {2024}),\ \bibinfo {note} {arXiv:2404.07020 [astro-ph]}\BibitemShut {NoStop}%
\bibitem [{\citenamefont {Agazie}\ \emph {et~al.}(2023{\natexlab{b}})\citenamefont {Agazie}, \citenamefont {Anumarlapudi}, \citenamefont {Archibald}, \citenamefont {Arzoumanian}, \citenamefont {Baker}, \citenamefont {Bécsy}, \citenamefont {Blecha}, \citenamefont {Brazier}, \citenamefont {Brook}, \citenamefont {Burke-Spolaor}, \citenamefont {Case}, \citenamefont {Casey-Clyde}, \citenamefont {Charisi}, \citenamefont {Chatterjee}, \citenamefont {Cohen}, \citenamefont {Cordes}, \citenamefont {Cornish}, \citenamefont {Crawford}, \citenamefont {Cromartie}, \citenamefont {Crowter}, \citenamefont {DeCesar}, \citenamefont {Demorest}, \citenamefont {Digman}, \citenamefont {Dolch}, \citenamefont {Drachler}, \citenamefont {Ferrara}, \citenamefont {Fiore}, \citenamefont {Fonseca}, \citenamefont {Freedman}, \citenamefont {Garver-Daniels}, \citenamefont {Gentile}, \citenamefont {Glaser}, \citenamefont {Good}, \citenamefont {Gültekin}, \citenamefont {Hazboun}, \citenamefont {Hourihane}, \citenamefont {Jennings},
  \citenamefont {Johnson}, \citenamefont {Jones}, \citenamefont {Kaiser}, \citenamefont {Kaplan}, \citenamefont {Kelley}, \citenamefont {Kerr}, \citenamefont {Key}, \citenamefont {Laal}, \citenamefont {Lam}, \citenamefont {Lamb}, \citenamefont {Lazio}, \citenamefont {Lewandowska}, \citenamefont {Liu}, \citenamefont {Lorimer}, \citenamefont {Luo}, \citenamefont {Lynch}, \citenamefont {Ma}, \citenamefont {Madison}, \citenamefont {McEwen}, \citenamefont {McKee}, \citenamefont {McLaughlin}, \citenamefont {McMann}, \citenamefont {Meyers}, \citenamefont {Meyers}, \citenamefont {Mingarelli}, \citenamefont {mitridate}, \citenamefont {natarajan}, \citenamefont {Ng}, \citenamefont {Nice}, \citenamefont {Ocker}, \citenamefont {Olum}, \citenamefont {Pennucci}, \citenamefont {Perera}, \citenamefont {Petrov}, \citenamefont {Pol}, \citenamefont {Radovan}, \citenamefont {Ransom}, \citenamefont {Ray}, \citenamefont {Romano}, \citenamefont {Sardesai}, \citenamefont {Schmiedekamp}, \citenamefont {Schmiedekamp}, \citenamefont
  {Schmitz}, \citenamefont {Shapiro-Albert}, \citenamefont {Siemens}, \citenamefont {Simon}, \citenamefont {Siwek}, \citenamefont {Stairs}, \citenamefont {Stinebring}, \citenamefont {Stovall}, \citenamefont {Susobhanan}, \citenamefont {Swiggum}, \citenamefont {Taylor}, \citenamefont {Taylor}, \citenamefont {Turner}, \citenamefont {Unal}, \citenamefont {Vallisneri}, \citenamefont {van Haasteren}, \citenamefont {Vigeland}, \citenamefont {Wahl}, \citenamefont {Witt},\ and\ \citenamefont {Young}}]{agazie_2023_NANOGrav15yearData_coherentsearch}%
  \BibitemOpen
  \bibfield  {author} {\bibinfo {author} {\bibfnamefont {G.}~\bibnamefont {Agazie}}, \bibinfo {author} {\bibfnamefont {A.}~\bibnamefont {Anumarlapudi}}, \bibinfo {author} {\bibfnamefont {A.~M.}\ \bibnamefont {Archibald}}, \emph {et~al.},\ }\href {https://doi.org/10.3847/2041-8213/ace18a} {\bibfield  {journal} {\bibinfo  {journal} {The Astrophysical Journal Letters}\ }\textbf {\bibinfo {volume} {951}},\ \bibinfo {pages} {L50} (\bibinfo {year} {2023}{\natexlab{b}})},\ \bibinfo {note} {arXiv:2306.16222 [astro-ph, physics:gr-qc]}\BibitemShut {NoStop}%
\bibitem [{\citenamefont {Amaro-Seoane}\ \emph {et~al.}(2017)\citenamefont {Amaro-Seoane}, \citenamefont {Audley}, \citenamefont {Babak}, \citenamefont {Baker}, \citenamefont {Barausse}, \citenamefont {Bender}, \citenamefont {Berti}, \citenamefont {Binetruy}, \citenamefont {Born}, \citenamefont {Bortoluzzi}, \citenamefont {Camp}, \citenamefont {Caprini}, \citenamefont {Cardoso}, \citenamefont {Colpi}, \citenamefont {Conklin}, \citenamefont {Cornish}, \citenamefont {Cutler}, \citenamefont {Danzmann}, \citenamefont {Dolesi}, \citenamefont {Ferraioli}, \citenamefont {Ferroni}, \citenamefont {Fitzsimons}, \citenamefont {Gair}, \citenamefont {Bote}, \citenamefont {Giardini}, \citenamefont {Gibert}, \citenamefont {Grimani}, \citenamefont {Halloin}, \citenamefont {Heinzel}, \citenamefont {Hertog}, \citenamefont {Hewitson}, \citenamefont {Holley-Bockelmann}, \citenamefont {Hollington}, \citenamefont {Hueller}, \citenamefont {Inchauspe}, \citenamefont {Jetzer}, \citenamefont {Karnesis}, \citenamefont {Killow},
  \citenamefont {Klein}, \citenamefont {Klipstein}, \citenamefont {Korsakova}, \citenamefont {Larson}, \citenamefont {Livas}, \citenamefont {Lloro}, \citenamefont {Man}, \citenamefont {Mance}, \citenamefont {Martino}, \citenamefont {Mateos}, \citenamefont {McKenzie}, \citenamefont {McWilliams}, \citenamefont {Miller}, \citenamefont {Mueller}, \citenamefont {Nardini}, \citenamefont {Nelemans}, \citenamefont {Nofrarias}, \citenamefont {Petiteau}, \citenamefont {Pivato}, \citenamefont {Plagnol}, \citenamefont {Porter}, \citenamefont {Reiche}, \citenamefont {Robertson}, \citenamefont {Robertson}, \citenamefont {Rossi}, \citenamefont {Russano}, \citenamefont {Schutz}, \citenamefont {Sesana}, \citenamefont {Shoemaker}, \citenamefont {Slutsky}, \citenamefont {Sopuerta}, \citenamefont {Sumner}, \citenamefont {Tamanini}, \citenamefont {Thorpe}, \citenamefont {Troebs}, \citenamefont {Vallisneri}, \citenamefont {Vecchio}, \citenamefont {Vetrugno}, \citenamefont {Vitale}, \citenamefont {Volonteri}, \citenamefont
  {Wanner}, \citenamefont {Ward}, \citenamefont {Wass}, \citenamefont {Weber}, \citenamefont {Ziemer},\ and\ \citenamefont {Zweifel}}]{amaro-seoane_2017_LaserInterferometerSpace}%
  \BibitemOpen
  \bibfield  {author} {\bibinfo {author} {\bibfnamefont {P.}~\bibnamefont {Amaro-Seoane}}, \bibinfo {author} {\bibfnamefont {H.}~\bibnamefont {Audley}}, \bibinfo {author} {\bibfnamefont {S.}~\bibnamefont {Babak}}, \emph {et~al.},\ }\href {http://arxiv.org/abs/1702.00786} {\bibinfo {title} {Laser {Interferometer} {Space} {Antenna}}} (\bibinfo {year} {2017}),\ \bibinfo {note} {arXiv:1702.00786 [astro-ph]}\BibitemShut {NoStop}%
\bibitem [{\citenamefont {Colpi}\ \emph {et~al.}(2024)\citenamefont {Colpi}, \citenamefont {Danzmann}, \citenamefont {Hewitson}, \citenamefont {Holley-Bockelmann}, \citenamefont {Jetzer}, \citenamefont {Nelemans}, \citenamefont {Petiteau}, \citenamefont {Shoemaker}, \citenamefont {Sopuerta}, \citenamefont {Stebbins}, \citenamefont {Tanvir}, \citenamefont {Ward}, \citenamefont {Weber}, \citenamefont {Thorpe}, \citenamefont {Daurskikh}, \citenamefont {Deep}, \citenamefont {Núñez}, \citenamefont {Marirrodriga}, \citenamefont {Gehler}, \citenamefont {Halain}, \citenamefont {Jennrich}, \citenamefont {Lammers}, \citenamefont {Larrañaga}, \citenamefont {Lieser}, \citenamefont {Lützgendorf}, \citenamefont {Martens}, \citenamefont {Mondin}, \citenamefont {Niño}, \citenamefont {Amaro-Seoane}, \citenamefont {Sedda}, \citenamefont {Auclair}, \citenamefont {Babak}, \citenamefont {Baghi}, \citenamefont {Baibhav}, \citenamefont {Baker}, \citenamefont {Bayle}, \citenamefont {Berry}, \citenamefont {Berti}, \citenamefont
  {Boileau}, \citenamefont {Bonetti}, \citenamefont {Brito}, \citenamefont {Buscicchio}, \citenamefont {Calcagni}, \citenamefont {Capelo}, \citenamefont {Caprini}, \citenamefont {Caputo}, \citenamefont {Castelli}, \citenamefont {Chen}, \citenamefont {Chen}, \citenamefont {Chua}, \citenamefont {Davies}, \citenamefont {Derdzinski}, \citenamefont {Domcke}, \citenamefont {Doneva}, \citenamefont {Dvorkin}, \citenamefont {Ezquiaga}, \citenamefont {Gair}, \citenamefont {Haiman}, \citenamefont {Harry}, \citenamefont {Hartwig}, \citenamefont {Hees}, \citenamefont {Heffernan}, \citenamefont {Husa}, \citenamefont {Izquierdo}, \citenamefont {Karnesis}, \citenamefont {Klein}, \citenamefont {Korol}, \citenamefont {Korsakova}, \citenamefont {Kupfer}, \citenamefont {Laghi}, \citenamefont {Lamberts}, \citenamefont {Larson}, \citenamefont {Jeune}, \citenamefont {Lewicki}, \citenamefont {Littenberg}, \citenamefont {Madge}, \citenamefont {Mangiagli}, \citenamefont {Marsat}, \citenamefont {Vilchez}, \citenamefont {Maselli},
  \citenamefont {Mathews}, \citenamefont {van~de Meent}, \citenamefont {Muratore}, \citenamefont {Nardini}, \citenamefont {Pani}, \citenamefont {Peloso}, \citenamefont {Pieroni}, \citenamefont {Pound}, \citenamefont {Quelquejay-Leclere}, \citenamefont {Ricciardone}, \citenamefont {Rossi}, \citenamefont {Sartirana}, \citenamefont {Savalle}, \citenamefont {Sberna}, \citenamefont {Sesana}, \citenamefont {Shoemaker}, \citenamefont {Slutsky}, \citenamefont {Sotiriou}, \citenamefont {Speri}, \citenamefont {Staab}, \citenamefont {Steer}, \citenamefont {Tamanini}, \citenamefont {Tasinato}, \citenamefont {Torrado}, \citenamefont {Torres-Orjuela}, \citenamefont {Toubiana}, \citenamefont {Vallisneri}, \citenamefont {Vecchio}, \citenamefont {Volonteri}, \citenamefont {Yagi},\ and\ \citenamefont {Zwick}}]{colpi_2024_LISADefinitionStudy}%
  \BibitemOpen
  \bibfield  {author} {\bibinfo {author} {\bibfnamefont {M.}~\bibnamefont {Colpi}}, \bibinfo {author} {\bibfnamefont {K.}~\bibnamefont {Danzmann}}, \bibinfo {author} {\bibfnamefont {M.}~\bibnamefont {Hewitson}}, \emph {et~al.},\ }\href {http://arxiv.org/abs/2402.07571} {\bibinfo {title} {{LISA} {Definition} {Study} {Report}}} (\bibinfo {year} {2024}),\ \bibinfo {note} {arXiv:2402.07571 [astro-ph, physics:gr-qc]}\BibitemShut {NoStop}%
\bibitem [{\citenamefont {Chang}\ and\ \citenamefont {Cui}(2020)}]{chang_2020_StochasticGravitationalWave}%
  \BibitemOpen
  \bibfield  {author} {\bibinfo {author} {\bibfnamefont {C.-F.}\ \bibnamefont {Chang}}\ and\ \bibinfo {author} {\bibfnamefont {Y.}~\bibnamefont {Cui}},\ }\href {https://doi.org/10.1016/j.dark.2020.100604} {\bibfield  {journal} {\bibinfo  {journal} {Physics of the Dark Universe}\ }\textbf {\bibinfo {volume} {29}},\ \bibinfo {pages} {100604} (\bibinfo {year} {2020})}\BibitemShut {NoStop}%
\bibitem [{\citenamefont {Domènech}\ and\ \citenamefont {Sasaki}(2024)}]{domenech_2024_ProbingPrimordialBlack}%
  \BibitemOpen
  \bibfield  {author} {\bibinfo {author} {\bibfnamefont {G.}~\bibnamefont {Domènech}}\ and\ \bibinfo {author} {\bibfnamefont {M.}~\bibnamefont {Sasaki}},\ }\href {https://doi.org/10.1088/1361-6382/ad5488} {\bibfield  {journal} {\bibinfo  {journal} {Classical and Quantum Gravity}\ }\textbf {\bibinfo {volume} {41}},\ \bibinfo {pages} {143001} (\bibinfo {year} {2024})}\BibitemShut {NoStop}%
\bibitem [{\citenamefont {Pyne}\ \emph {et~al.}(1996)\citenamefont {Pyne}, \citenamefont {Gwinn}, \citenamefont {Birkinshaw}, \citenamefont {Eubanks},\ and\ \citenamefont {Matsakis}}]{pyne_1996_GravitationalRadiationVerya}%
  \BibitemOpen
  \bibfield  {author} {\bibinfo {author} {\bibfnamefont {T.}~\bibnamefont {Pyne}}, \bibinfo {author} {\bibfnamefont {C.~R.}\ \bibnamefont {Gwinn}}, \bibinfo {author} {\bibfnamefont {M.}~\bibnamefont {Birkinshaw}}, \emph {et~al.},\ }\href {https://doi.org/10.1086/177443} {\bibfield  {journal} {\bibinfo  {journal} {The Astrophysical Journal}\ }\textbf {\bibinfo {volume} {465}},\ \bibinfo {pages} {566} (\bibinfo {year} {1996})}\BibitemShut {NoStop}%
\bibitem [{\citenamefont {Book}\ and\ \citenamefont {Flanagan}(2011)}]{book_2011_AstrometricEffectsStochastica}%
  \BibitemOpen
  \bibfield  {author} {\bibinfo {author} {\bibfnamefont {L.~G.}\ \bibnamefont {Book}}\ and\ \bibinfo {author} {\bibfnamefont {E.~E.}\ \bibnamefont {Flanagan}},\ }\href {https://doi.org/10.1103/PhysRevD.83.024024} {\bibfield  {journal} {\bibinfo  {journal} {Physical Review D}\ }\textbf {\bibinfo {volume} {83}},\ \bibinfo {pages} {024024} (\bibinfo {year} {2011})}\BibitemShut {NoStop}%
\bibitem [{\citenamefont {Klioner}(2018)}]{klioner_2018_GaialikeAstrometryGravitational}%
  \BibitemOpen
  \bibfield  {author} {\bibinfo {author} {\bibfnamefont {S.~A.}\ \bibnamefont {Klioner}},\ }\href {https://doi.org/10.1088/1361-6382/aa9f57} {\bibfield  {journal} {\bibinfo  {journal} {Classical and Quantum Gravity}\ }\textbf {\bibinfo {volume} {35}},\ \bibinfo {pages} {045005} (\bibinfo {year} {2018})}\BibitemShut {NoStop}%
\bibitem [{\citenamefont {Moore}\ \emph {et~al.}(2017)\citenamefont {Moore}, \citenamefont {Mihaylov}, \citenamefont {Lasenby},\ and\ \citenamefont {Gilmore}}]{moore_2017_gaia_compression}%
  \BibitemOpen
  \bibfield  {author} {\bibinfo {author} {\bibfnamefont {C.~J.}\ \bibnamefont {Moore}}, \bibinfo {author} {\bibfnamefont {D.~P.}\ \bibnamefont {Mihaylov}}, \bibinfo {author} {\bibfnamefont {A.}~\bibnamefont {Lasenby}},\ and\ \bibinfo {author} {\bibfnamefont {G.}~\bibnamefont {Gilmore}},\ }\href {https://doi.org/10.1103/PhysRevLett.119.261102} {\bibfield  {journal} {\bibinfo  {journal} {Physical Review Letters}\ }\textbf {\bibinfo {volume} {119}},\ \bibinfo {pages} {261102} (\bibinfo {year} {2017})}\BibitemShut {NoStop}%
\bibitem [{\citenamefont {{Wang}}\ \emph {et~al.}(2021)\citenamefont {{Wang}}, \citenamefont {{Pardo}}, \citenamefont {{Chang}},\ and\ \citenamefont {{Dor{\'e}}}}]{wang+21_romangw_coherent}%
  \BibitemOpen
  \bibfield  {author} {\bibinfo {author} {\bibfnamefont {Y.}~\bibnamefont {{Wang}}}, \bibinfo {author} {\bibfnamefont {K.}~\bibnamefont {{Pardo}}}, \bibinfo {author} {\bibfnamefont {T.-C.}\ \bibnamefont {{Chang}}},\ and\ \bibinfo {author} {\bibfnamefont {O.}~\bibnamefont {{Dor{\'e}}}},\ }\href {https://doi.org/10.1103/PhysRevD.103.084007} {\bibfield  {journal} {\bibinfo  {journal} {\prd}\ }\textbf {\bibinfo {volume} {103}},\ \bibinfo {eid} {084007} (\bibinfo {year} {2021})},\ \Eprint {https://arxiv.org/abs/2010.02218} {arXiv:2010.02218 [gr-qc]} \BibitemShut {NoStop}%
\bibitem [{\citenamefont {{Wang}}\ \emph {et~al.}(2022)\citenamefont {{Wang}}, \citenamefont {{Pardo}}, \citenamefont {{Chang}},\ and\ \citenamefont {{Dor{\'e}}}}]{wang+22_romangw_background}%
  \BibitemOpen
  \bibfield  {author} {\bibinfo {author} {\bibfnamefont {Y.}~\bibnamefont {{Wang}}}, \bibinfo {author} {\bibfnamefont {K.}~\bibnamefont {{Pardo}}}, \bibinfo {author} {\bibfnamefont {T.-C.}\ \bibnamefont {{Chang}}},\ and\ \bibinfo {author} {\bibfnamefont {O.}~\bibnamefont {{Dor{\'e}}}},\ }\href {https://doi.org/10.1103/PhysRevD.106.084006} {\bibfield  {journal} {\bibinfo  {journal} {\prd}\ }\textbf {\bibinfo {volume} {106}},\ \bibinfo {eid} {084006} (\bibinfo {year} {2022})},\ \Eprint {https://arxiv.org/abs/2205.07962} {arXiv:2205.07962 [gr-qc]} \BibitemShut {NoStop}%
\bibitem [{\citenamefont {Pardo}\ \emph {et~al.}(2023)\citenamefont {Pardo}, \citenamefont {Chang}, \citenamefont {Doré},\ and\ \citenamefont {Wang}}]{pardo_2023_GravitationalWaveDetectiona}%
  \BibitemOpen
  \bibfield  {author} {\bibinfo {author} {\bibfnamefont {K.}~\bibnamefont {Pardo}}, \bibinfo {author} {\bibfnamefont {T.-C.}\ \bibnamefont {Chang}}, \bibinfo {author} {\bibfnamefont {O.}~\bibnamefont {Doré}},\ and\ \bibinfo {author} {\bibfnamefont {Y.}~\bibnamefont {Wang}},\ }\href {https://doi.org/10.48550/ARXIV.2306.14968} {\bibinfo {title} {Gravitational {Wave} {Detection} with {Relative} {Astrometry} using {Roman}'s {Galactic} {Bulge} {Time} {Domain} {Survey}}} (\bibinfo {year} {2023}),\ \bibinfo {note} {version Number: 1}\BibitemShut {NoStop}%
\bibitem [{\citenamefont {{B{\'e}csy}}\ \emph {et~al.}(2022)\citenamefont {{B{\'e}csy}}, \citenamefont {{Cornish}},\ and\ \citenamefont {{Digman}}}]{becsy+22_quickcw1}%
  \BibitemOpen
  \bibfield  {author} {\bibinfo {author} {\bibfnamefont {B.}~\bibnamefont {{B{\'e}csy}}}, \bibinfo {author} {\bibfnamefont {N.~J.}\ \bibnamefont {{Cornish}}},\ and\ \bibinfo {author} {\bibfnamefont {M.~C.}\ \bibnamefont {{Digman}}},\ }\href {https://doi.org/10.1103/PhysRevD.105.122003} {\bibfield  {journal} {\bibinfo  {journal} {\prd}\ }\textbf {\bibinfo {volume} {105}},\ \bibinfo {eid} {122003} (\bibinfo {year} {2022})},\ \Eprint {https://arxiv.org/abs/2204.07160} {arXiv:2204.07160 [gr-qc]} \BibitemShut {NoStop}%
\bibitem [{\citenamefont {{B{\'e}csy}}(2024)}]{becsy+24_quickcw2}%
  \BibitemOpen
  \bibfield  {author} {\bibinfo {author} {\bibfnamefont {B.}~\bibnamefont {{B{\'e}csy}}},\ }\href {https://doi.org/10.48550/arXiv.2406.16331} {\bibfield  {journal} {\bibinfo  {journal} {arXiv e-prints}\ ,\ \bibinfo {eid} {arXiv:2406.16331}} (\bibinfo {year} {2024})},\ \Eprint {https://arxiv.org/abs/2406.16331} {arXiv:2406.16331 [gr-qc]} \BibitemShut {NoStop}%
\bibitem [{\citenamefont {{Planck Collaboration}}\ \emph {et~al.}(2020)\citenamefont {{Planck Collaboration}}, \citenamefont {Aghanim}, \citenamefont {Akrami}, \citenamefont {Ashdown}, \citenamefont {Aumont}, \citenamefont {Baccigalupi}, \citenamefont {Ballardini}, \citenamefont {Banday}, \citenamefont {Barreiro}, \citenamefont {Bartolo}, \citenamefont {Basak}, \citenamefont {Battye}, \citenamefont {Benabed}, \citenamefont {Bernard}, \citenamefont {Bersanelli}, \citenamefont {Bielewicz}, \citenamefont {Bock}, \citenamefont {Bond}, \citenamefont {Borrill}, \citenamefont {Bouchet}, \citenamefont {Boulanger}, \citenamefont {Bucher}, \citenamefont {Burigana}, \citenamefont {Butler}, \citenamefont {Calabrese}, \citenamefont {Cardoso}, \citenamefont {Carron}, \citenamefont {Challinor}, \citenamefont {Chiang}, \citenamefont {Chluba}, \citenamefont {Colombo}, \citenamefont {Combet}, \citenamefont {Contreras}, \citenamefont {Crill}, \citenamefont {Cuttaia}, \citenamefont {De~Bernardis}, \citenamefont {De~Zotti},
  \citenamefont {Delabrouille}, \citenamefont {Delouis}, \citenamefont {Di~Valentino}, \citenamefont {Diego}, \citenamefont {Doré}, \citenamefont {Douspis}, \citenamefont {Ducout}, \citenamefont {Dupac}, \citenamefont {Dusini}, \citenamefont {Efstathiou}, \citenamefont {Elsner}, \citenamefont {Enßlin}, \citenamefont {Eriksen}, \citenamefont {Fantaye}, \citenamefont {Farhang}, \citenamefont {Fergusson}, \citenamefont {Fernandez-Cobos}, \citenamefont {Finelli}, \citenamefont {Forastieri}, \citenamefont {Frailis}, \citenamefont {Fraisse}, \citenamefont {Franceschi}, \citenamefont {Frolov}, \citenamefont {Galeotta}, \citenamefont {Galli}, \citenamefont {Ganga}, \citenamefont {Génova-Santos}, \citenamefont {Gerbino}, \citenamefont {Ghosh}, \citenamefont {González-Nuevo}, \citenamefont {Górski}, \citenamefont {Gratton}, \citenamefont {Gruppuso}, \citenamefont {Gudmundsson}, \citenamefont {Hamann}, \citenamefont {Handley}, \citenamefont {Hansen}, \citenamefont {Herranz}, \citenamefont {Hildebrandt},
  \citenamefont {Hivon}, \citenamefont {Huang}, \citenamefont {Jaffe}, \citenamefont {Jones}, \citenamefont {Karakci}, \citenamefont {Keihänen}, \citenamefont {Keskitalo}, \citenamefont {Kiiveri}, \citenamefont {Kim}, \citenamefont {Kisner}, \citenamefont {Knox}, \citenamefont {Krachmalnicoff}, \citenamefont {Kunz}, \citenamefont {Kurki-Suonio}, \citenamefont {Lagache}, \citenamefont {Lamarre}, \citenamefont {Lasenby}, \citenamefont {Lattanzi}, \citenamefont {Lawrence}, \citenamefont {Le~Jeune}, \citenamefont {Lemos}, \citenamefont {Lesgourgues}, \citenamefont {Levrier}, \citenamefont {Lewis}, \citenamefont {Liguori}, \citenamefont {Lilje}, \citenamefont {Lilley}, \citenamefont {Lindholm}, \citenamefont {López-Caniego}, \citenamefont {Lubin}, \citenamefont {Ma}, \citenamefont {Macías-Pérez}, \citenamefont {Maggio}, \citenamefont {Maino}, \citenamefont {Mandolesi}, \citenamefont {Mangilli}, \citenamefont {Marcos-Caballero}, \citenamefont {Maris}, \citenamefont {Martin}, \citenamefont {Martinelli},
  \citenamefont {Martínez-González}, \citenamefont {Matarrese}, \citenamefont {Mauri}, \citenamefont {McEwen}, \citenamefont {Meinhold}, \citenamefont {Melchiorri}, \citenamefont {Mennella}, \citenamefont {Migliaccio}, \citenamefont {Millea}, \citenamefont {Mitra}, \citenamefont {Miville-Deschênes}, \citenamefont {Molinari}, \citenamefont {Montier}, \citenamefont {Morgante}, \citenamefont {Moss}, \citenamefont {Natoli}, \citenamefont {Nørgaard-Nielsen}, \citenamefont {Pagano}, \citenamefont {Paoletti}, \citenamefont {Partridge}, \citenamefont {Patanchon}, \citenamefont {Peiris}, \citenamefont {Perrotta}, \citenamefont {Pettorino}, \citenamefont {Piacentini}, \citenamefont {Polastri}, \citenamefont {Polenta}, \citenamefont {Puget}, \citenamefont {Rachen}, \citenamefont {Reinecke}, \citenamefont {Remazeilles}, \citenamefont {Renzi}, \citenamefont {Rocha}, \citenamefont {Rosset}, \citenamefont {Roudier}, \citenamefont {Rubiño-Martín}, \citenamefont {Ruiz-Granados}, \citenamefont {Salvati}, \citenamefont
  {Sandri}, \citenamefont {Savelainen}, \citenamefont {Scott}, \citenamefont {Shellard}, \citenamefont {Sirignano}, \citenamefont {Sirri}, \citenamefont {Spencer}, \citenamefont {Sunyaev}, \citenamefont {Suur-Uski}, \citenamefont {Tauber}, \citenamefont {Tavagnacco}, \citenamefont {Tenti}, \citenamefont {Toffolatti}, \citenamefont {Tomasi}, \citenamefont {Trombetti}, \citenamefont {Valenziano}, \citenamefont {Valiviita}, \citenamefont {Van~Tent}, \citenamefont {Vibert}, \citenamefont {Vielva}, \citenamefont {Villa}, \citenamefont {Vittorio}, \citenamefont {Wandelt}, \citenamefont {Wehus}, \citenamefont {White}, \citenamefont {White}, \citenamefont {Zacchei},\ and\ \citenamefont {Zonca}}]{planckcollaboration_2020_Planck2018Results}%
  \BibitemOpen
  \bibfield  {author} {\bibinfo {author} {\bibnamefont {{Planck Collaboration}}}, \bibinfo {author} {\bibfnamefont {N.}~\bibnamefont {Aghanim}}, \bibinfo {author} {\bibfnamefont {Y.}~\bibnamefont {Akrami}}, \emph {et~al.},\ }\href {https://doi.org/10.1051/0004-6361/201833910} {\bibfield  {journal} {\bibinfo  {journal} {Astronomy \& Astrophysics}\ }\textbf {\bibinfo {volume} {641}},\ \bibinfo {pages} {A6} (\bibinfo {year} {2020})}\BibitemShut {NoStop}%
\bibitem [{\citenamefont {Gaudi}\ and\ \citenamefont {Bennett}(2023)}]{gaudi_2023_RomanGalacticExoplanet}%
  \BibitemOpen
  \bibfield  {author} {\bibinfo {author} {\bibfnamefont {B.~S.}\ \bibnamefont {Gaudi}}\ and\ \bibinfo {author} {\bibfnamefont {D.~P.}\ \bibnamefont {Bennett}},\ }\href {https://asd.gsfc.nasa.gov/roman/wps_2023/files/041_Bennett_GBTDS.pdf} {\bibfield  {journal} {\bibinfo  {journal} {Roman White Papers - 2023}\ } (\bibinfo {year} {2023})}\BibitemShut {NoStop}%
\bibitem [{\citenamefont {Gaudi}\ \emph {et~al.}(2019)\citenamefont {Gaudi}, \citenamefont {Akeson}, \citenamefont {Anderson}, \citenamefont {Bachelet}, \citenamefont {Bennett}, \citenamefont {Bhattacharya}, \citenamefont {Bozza}, \citenamefont {Novati}, \citenamefont {Henderson}, \citenamefont {Johnson}, \citenamefont {Kruk}, \citenamefont {Lu}, \citenamefont {Mao}, \citenamefont {Montet}, \citenamefont {Nataf}, \citenamefont {Penny}, \citenamefont {Poleski}, \citenamefont {Ranc}, \citenamefont {Sahu}, \citenamefont {Shvartzvald}, \citenamefont {Spergel}, \citenamefont {Suzuki}, \citenamefont {Stassun},\ and\ \citenamefont {Street}}]{gaudi_2019_AuxiliaryScienceWFIRST}%
  \BibitemOpen
  \bibfield  {author} {\bibinfo {author} {\bibfnamefont {B.~S.}\ \bibnamefont {Gaudi}}, \bibinfo {author} {\bibfnamefont {R.}~\bibnamefont {Akeson}}, \bibinfo {author} {\bibfnamefont {J.}~\bibnamefont {Anderson}}, \emph {et~al.},\ }\href {https://doi.org/10.48550/ARXIV.1903.08986} {\bibinfo {title} {"{Auxiliary}" {Science} with the {WFIRST} {Microlensing} {Survey}}} (\bibinfo {year} {2019}),\ \bibinfo {note} {version Number: 1}\BibitemShut {NoStop}%
\bibitem [{\citenamefont {{Borucki}}\ \emph {et~al.}(2010)\citenamefont {{Borucki}}, \citenamefont {{Koch}}, \citenamefont {{Basri}}, \citenamefont {{Batalha}}, \citenamefont {{Brown}}, \citenamefont {{Caldwell}}, \citenamefont {{Caldwell}}, \citenamefont {{Christensen-Dalsgaard}}, \citenamefont {{Cochran}}, \citenamefont {{DeVore}}, \citenamefont {{Dunham}}, \citenamefont {{Dupree}}, \citenamefont {{Gautier}}, \citenamefont {{Geary}}, \citenamefont {{Gilliland}}, \citenamefont {{Gould}}, \citenamefont {{Howell}}, \citenamefont {{Jenkins}}, \citenamefont {{Kondo}}, \citenamefont {{Latham}}, \citenamefont {{Marcy}}, \citenamefont {{Meibom}}, \citenamefont {{Kjeldsen}}, \citenamefont {{Lissauer}}, \citenamefont {{Monet}}, \citenamefont {{Morrison}}, \citenamefont {{Sasselov}}, \citenamefont {{Tarter}}, \citenamefont {{Boss}}, \citenamefont {{Brownlee}}, \citenamefont {{Owen}}, \citenamefont {{Buzasi}}, \citenamefont {{Charbonneau}}, \citenamefont {{Doyle}}, \citenamefont {{Fortney}}, \citenamefont {{Ford}},
  \citenamefont {{Holman}}, \citenamefont {{Seager}}, \citenamefont {{Steffen}}, \citenamefont {{Welsh}}, \citenamefont {{Rowe}}, \citenamefont {{Anderson}}, \citenamefont {{Buchhave}}, \citenamefont {{Ciardi}}, \citenamefont {{Walkowicz}}, \citenamefont {{Sherry}}, \citenamefont {{Horch}}, \citenamefont {{Isaacson}}, \citenamefont {{Everett}}, \citenamefont {{Fischer}}, \citenamefont {{Torres}}, \citenamefont {{Johnson}}, \citenamefont {{Endl}}, \citenamefont {{MacQueen}}, \citenamefont {{Bryson}}, \citenamefont {{Dotson}}, \citenamefont {{Haas}}, \citenamefont {{Kolodziejczak}}, \citenamefont {{Van Cleve}}, \citenamefont {{Chandrasekaran}}, \citenamefont {{Twicken}}, \citenamefont {{Quintana}}, \citenamefont {{Clarke}}, \citenamefont {{Allen}}, \citenamefont {{Li}}, \citenamefont {{Wu}}, \citenamefont {{Tenenbaum}}, \citenamefont {{Verner}}, \citenamefont {{Bruhweiler}}, \citenamefont {{Barnes}},\ and\ \citenamefont {{Prsa}}}]{Borucki2010}%
  \BibitemOpen
  \bibfield  {author} {\bibinfo {author} {\bibfnamefont {W.~J.}\ \bibnamefont {{Borucki}}}, \bibinfo {author} {\bibfnamefont {D.}~\bibnamefont {{Koch}}}, \bibinfo {author} {\bibfnamefont {G.}~\bibnamefont {{Basri}}}, \emph {et~al.},\ }\href {https://doi.org/10.1126/science.1185402} {\bibfield  {journal} {\bibinfo  {journal} {Science}\ }\textbf {\bibinfo {volume} {327}},\ \bibinfo {pages} {977} (\bibinfo {year} {2010})}\BibitemShut {NoStop}%
\bibitem [{\citenamefont {{Van Cleve}}\ \emph {et~al.}(2016)\citenamefont {{Van Cleve}}, \citenamefont {{Christiansen}}, \citenamefont {{Jenkins}}, \citenamefont {{Caldwell}}, \citenamefont {{Barclay}}, \citenamefont {{Bryson}}, \citenamefont {{Burke}}, \citenamefont {{Cambell}}, \citenamefont {{Catanzarite}}, \citenamefont {{Clarke}}, \citenamefont {{Coughlin}}, \citenamefont {{Girouard}}, \citenamefont {{Haas}}, \citenamefont {{Klaus}}, \citenamefont {{Kolodziejczak}}, \citenamefont {{Li}}, \citenamefont {{McCauliff}}, \citenamefont {{Morris}}, \citenamefont {{Mullally}}, \citenamefont {{Quintana}}, \citenamefont {{Rowe}}, \citenamefont {{Sabale}}, \citenamefont {{Seader}}, \citenamefont {{Smith}}, \citenamefont {{Still}}, \citenamefont {{Tenenbaum}}, \citenamefont {{Thompson}}, \citenamefont {{Twicken}}, \citenamefont {{Kamal Uddin}},\ and\ \citenamefont {{Zamudio}}}]{VanCleve2016}%
  \BibitemOpen
  \bibfield  {author} {\bibinfo {author} {\bibfnamefont {J.~E.}\ \bibnamefont {{Van Cleve}}}, \bibinfo {author} {\bibfnamefont {J.~L.}\ \bibnamefont {{Christiansen}}}, \bibinfo {author} {\bibfnamefont {J.~M.}\ \bibnamefont {{Jenkins}}}, \emph {et~al.},\ }\href@noop {} {\bibinfo {title} {{Kepler Data Characteristics Handbook}}},\ \bibinfo {howpublished} {Kepler Science Document KSCI-19040-005, id. 2. Edited by Doug Caldwell, Jon M. Jenkins, Michael R. Haas and Natalie Batalha} (\bibinfo {year} {2016})\BibitemShut {NoStop}%
\bibitem [{\citenamefont {Monet}\ \emph {et~al.}(2010)\citenamefont {Monet}, \citenamefont {Jenkins}, \citenamefont {Dunham}, \citenamefont {Bryson}, \citenamefont {Gilliland}, \citenamefont {Latham}, \citenamefont {Borucki},\ and\ \citenamefont {Koch}}]{monet_2010_PreliminaryAstrometricResults}%
  \BibitemOpen
  \bibfield  {author} {\bibinfo {author} {\bibfnamefont {D.~G.}\ \bibnamefont {Monet}}, \bibinfo {author} {\bibfnamefont {J.~M.}\ \bibnamefont {Jenkins}}, \bibinfo {author} {\bibfnamefont {E.~W.}\ \bibnamefont {Dunham}}, \emph {et~al.},\ }\href {https://doi.org/10.48550/ARXIV.1001.0305} {\bibinfo {title} {Preliminary {Astrometric} {Results} from {Kepler}}} (\bibinfo {year} {2010}),\ \bibinfo {note} {version Number: 1}\BibitemShut {NoStop}%
\bibitem [{\citenamefont {{Brown}}\ \emph {et~al.}(2011)\citenamefont {{Brown}}, \citenamefont {{Latham}}, \citenamefont {{Everett}},\ and\ \citenamefont {{Esquerdo}}}]{brown_2011_kic}%
  \BibitemOpen
  \bibfield  {author} {\bibinfo {author} {\bibfnamefont {T.~M.}\ \bibnamefont {{Brown}}}, \bibinfo {author} {\bibfnamefont {D.~W.}\ \bibnamefont {{Latham}}}, \bibinfo {author} {\bibfnamefont {M.~E.}\ \bibnamefont {{Everett}}},\ and\ \bibinfo {author} {\bibfnamefont {G.~A.}\ \bibnamefont {{Esquerdo}}},\ }\href {https://doi.org/10.1088/0004-6256/142/4/112} {\bibfield  {journal} {\bibinfo  {journal} {\aj}\ }\textbf {\bibinfo {volume} {142}},\ \bibinfo {eid} {112} (\bibinfo {year} {2011})},\ \Eprint {https://arxiv.org/abs/1102.0342} {arXiv:1102.0342 [astro-ph.SR]} \BibitemShut {NoStop}%
\bibitem [{\citenamefont {{Gaia Collaboration}}\ \emph {et~al.}(2023)\citenamefont {{Gaia Collaboration}}, \citenamefont {{Vallenari}}, \citenamefont {{Brown}}, \citenamefont {{Prusti}}, \citenamefont {{de Bruijne}}, \citenamefont {{Arenou}}, \citenamefont {{Babusiaux}}, \citenamefont {{Biermann}}, \citenamefont {{Creevey}}, \citenamefont {{Ducourant}}, \citenamefont {{Evans}}, \citenamefont {{Eyer}}, \citenamefont {{Guerra}}, \citenamefont {{Hutton}}, \citenamefont {{Jordi}}, \citenamefont {{Klioner}}, \citenamefont {{Lammers}}, \citenamefont {{Lindegren}}, \citenamefont {{Luri}}, \citenamefont {{Mignard}}, \citenamefont {{Panem}}, \citenamefont {{Pourbaix}}, \citenamefont {{Randich}}, \citenamefont {{Sartoretti}}, \citenamefont {{Soubiran}}, \citenamefont {{Tanga}}, \citenamefont {{Walton}}, \citenamefont {{Bailer-Jones}}, \citenamefont {{Bastian}}, \citenamefont {{Drimmel}}, \citenamefont {{Jansen}}, \citenamefont {{Katz}}, \citenamefont {{Lattanzi}}, \citenamefont {{van Leeuwen}}, \citenamefont
  {{Bakker}}, \citenamefont {{Cacciari}}, \citenamefont {{Casta{\~n}eda}}, \citenamefont {{De Angeli}}, \citenamefont {{Fabricius}}, \citenamefont {{Fouesneau}}, \citenamefont {{Fr{\'e}mat}}, \citenamefont {{Galluccio}}, \citenamefont {{Guerrier}}, \citenamefont {{Heiter}}, \citenamefont {{Masana}}, \citenamefont {{Messineo}}, \citenamefont {{Mowlavi}}, \citenamefont {{Nicolas}}, \citenamefont {{Nienartowicz}}, \citenamefont {{Pailler}}, \citenamefont {{Panuzzo}}, \citenamefont {{Riclet}}, \citenamefont {{Roux}}, \citenamefont {{Seabroke}}, \citenamefont {{Sordo}}, \citenamefont {{Th{\'e}venin}}, \citenamefont {{Gracia-Abril}}, \citenamefont {{Portell}}, \citenamefont {{Teyssier}}, \citenamefont {{Altmann}}, \citenamefont {{Andrae}}, \citenamefont {{Audard}}, \citenamefont {{Bellas-Velidis}}, \citenamefont {{Benson}}, \citenamefont {{Berthier}}, \citenamefont {{Blomme}}, \citenamefont {{Burgess}}, \citenamefont {{Busonero}}, \citenamefont {{Busso}}, \citenamefont {{C{\'a}novas}}, \citenamefont {{Carry}},
  \citenamefont {{Cellino}}, \citenamefont {{Cheek}}, \citenamefont {{Clementini}}, \citenamefont {{Damerdji}}, \citenamefont {{Davidson}}, \citenamefont {{de Teodoro}}, \citenamefont {{Nu{\~n}ez Campos}}, \citenamefont {{Delchambre}}, \citenamefont {{Dell'Oro}}, \citenamefont {{Esquej}}, \citenamefont {{Fern{\'a}ndez-Hern{\'a}ndez}}, \citenamefont {{Fraile}}, \citenamefont {{Garabato}}, \citenamefont {{Garc{\'\i}a-Lario}}, \citenamefont {{Gosset}}, \citenamefont {{Haigron}}, \citenamefont {{Halbwachs}}, \citenamefont {{Hambly}}, \citenamefont {{Harrison}}, \citenamefont {{Hern{\'a}ndez}}, \citenamefont {{Hestroffer}}, \citenamefont {{Hodgkin}}, \citenamefont {{Holl}}, \citenamefont {{Jan{\ss}en}}, \citenamefont {{Jevardat de Fombelle}}, \citenamefont {{Jordan}}, \citenamefont {{Krone-Martins}}, \citenamefont {{Lanzafame}}, \citenamefont {{L{\"o}ffler}}, \citenamefont {{Marchal}}, \citenamefont {{Marrese}}, \citenamefont {{Moitinho}}, \citenamefont {{Muinonen}}, \citenamefont {{Osborne}}, \citenamefont
  {{Pancino}}, \citenamefont {{Pauwels}}, \citenamefont {{Recio-Blanco}}, \citenamefont {{Reyl{\'e}}}, \citenamefont {{Riello}}, \citenamefont {{Rimoldini}}, \citenamefont {{Roegiers}}, \citenamefont {{Rybizki}}, \citenamefont {{Sarro}}, \citenamefont {{Siopis}}, \citenamefont {{Smith}}, \citenamefont {{Sozzetti}}, \citenamefont {{Utrilla}}, \citenamefont {{van Leeuwen}}, \citenamefont {{Abbas}}, \citenamefont {{{\'A}brah{\'a}m}}, \citenamefont {{Abreu Aramburu}}, \citenamefont {{Aerts}}, \citenamefont {{Aguado}}, \citenamefont {{Ajaj}}, \citenamefont {{Aldea-Montero}}, \citenamefont {{Altavilla}}, \citenamefont {{{\'A}lvarez}}, \citenamefont {{Alves}}, \citenamefont {{Anders}}, \citenamefont {{Anderson}}, \citenamefont {{Anglada Varela}}, \citenamefont {{Antoja}}, \citenamefont {{Baines}}, \citenamefont {{Baker}}, \citenamefont {{Balaguer-N{\'u}{\~n}ez}}, \citenamefont {{Balbinot}}, \citenamefont {{Balog}}, \citenamefont {{Barache}}, \citenamefont {{Barbato}}, \citenamefont {{Barros}}, \citenamefont
  {{Barstow}}, \citenamefont {{Bartolom{\'e}}}, \citenamefont {{Bassilana}}, \citenamefont {{Bauchet}}, \citenamefont {{Becciani}}, \citenamefont {{Bellazzini}}, \citenamefont {{Berihuete}}, \citenamefont {{Bernet}}, \citenamefont {{Bertone}}, \citenamefont {{Bianchi}}, \citenamefont {{Binnenfeld}}, \citenamefont {{Blanco-Cuaresma}}, \citenamefont {{Blazere}}, \citenamefont {{Boch}}, \citenamefont {{Bombrun}}, \citenamefont {{Bossini}}, \citenamefont {{Bouquillon}}, \citenamefont {{Bragaglia}}, \citenamefont {{Bramante}}, \citenamefont {{Breedt}}, \citenamefont {{Bressan}}, \citenamefont {{Brouillet}}, \citenamefont {{Brugaletta}}, \citenamefont {{Bucciarelli}}, \citenamefont {{Burlacu}}, \citenamefont {{Butkevich}}, \citenamefont {{Buzzi}}, \citenamefont {{Caffau}}, \citenamefont {{Cancelliere}}, \citenamefont {{Cantat-Gaudin}}, \citenamefont {{Carballo}}, \citenamefont {{Carlucci}}, \citenamefont {{Carnerero}}, \citenamefont {{Carrasco}}, \citenamefont {{Casamiquela}}, \citenamefont {{Castellani}},
  \citenamefont {{Castro-Ginard}}, \citenamefont {{Chaoul}}, \citenamefont {{Charlot}}, \citenamefont {{Chemin}}, \citenamefont {{Chiaramida}}, \citenamefont {{Chiavassa}}, \citenamefont {{Chornay}}, \citenamefont {{Comoretto}}, \citenamefont {{Contursi}}, \citenamefont {{Cooper}}, \citenamefont {{Cornez}}, \citenamefont {{Cowell}}, \citenamefont {{Crifo}}, \citenamefont {{Cropper}}, \citenamefont {{Crosta}}, \citenamefont {{Crowley}}, \citenamefont {{Dafonte}}, \citenamefont {{Dapergolas}}, \citenamefont {{David}}, \citenamefont {{David}}, \citenamefont {{de Laverny}}, \citenamefont {{De Luise}},\ and\ \citenamefont {{De March}}}]{gaia_2023_dr3}%
  \BibitemOpen
  \bibfield  {author} {\bibinfo {author} {\bibnamefont {{Gaia Collaboration}}}, \bibinfo {author} {\bibfnamefont {A.}~\bibnamefont {{Vallenari}}}, \bibinfo {author} {\bibfnamefont {A.~G.~A.}\ \bibnamefont {{Brown}}}, \emph {et~al.},\ }\href {https://doi.org/10.1051/0004-6361/202243940} {\bibfield  {journal} {\bibinfo  {journal} {\aap}\ }\textbf {\bibinfo {volume} {674}},\ \bibinfo {eid} {A1} (\bibinfo {year} {2023})},\ \Eprint {https://arxiv.org/abs/2208.00211} {arXiv:2208.00211 [astro-ph.GA]} \BibitemShut {NoStop}%
\bibitem [{\citenamefont {{Janes}}(2017)}]{janes_2017_kepler_commonpropermotions}%
  \BibitemOpen
  \bibfield  {author} {\bibinfo {author} {\bibfnamefont {K.~A.}\ \bibnamefont {{Janes}}},\ }\href {https://doi.org/10.3847/1538-4357/835/1/75} {\bibfield  {journal} {\bibinfo  {journal} {\apj}\ }\textbf {\bibinfo {volume} {835}},\ \bibinfo {eid} {75} (\bibinfo {year} {2017})},\ \Eprint {https://arxiv.org/abs/1612.00070} {arXiv:1612.00070 [astro-ph.SR]} \BibitemShut {NoStop}%
\bibitem [{\citenamefont {{WFIRST Astrometry Working Group}}\ \emph {et~al.}(2019)\citenamefont {{WFIRST Astrometry Working Group}}, \citenamefont {{Sanderson}}, \citenamefont {{Bellini}}, \citenamefont {{Casertano}}, \citenamefont {{Lu}}, \citenamefont {{Melchior}}, \citenamefont {{Libralato}}, \citenamefont {{Bennett}}, \citenamefont {{Shao}}, \citenamefont {{Rhodes}}, \citenamefont {{Sohn}}, \citenamefont {{Malhotra}}, \citenamefont {{Gaudi}}, \citenamefont {{Fall}}, \citenamefont {{Nelan}}, \citenamefont {{Guhathakurta}}, \citenamefont {{Anderson}},\ and\ \citenamefont {{Ho}}}]{wfirst+19_roman_astrometry}%
  \BibitemOpen
  \bibfield  {author} {\bibinfo {author} {\bibnamefont {{WFIRST Astrometry Working Group}}}, \bibinfo {author} {\bibfnamefont {R.~E.}\ \bibnamefont {{Sanderson}}}, \bibinfo {author} {\bibfnamefont {A.}~\bibnamefont {{Bellini}}}, \emph {et~al.},\ }\href {https://doi.org/10.1117/1.JATIS.5.4.044005} {\bibfield  {journal} {\bibinfo  {journal} {Journal of Astronomical Telescopes, Instruments, and Systems}\ }\textbf {\bibinfo {volume} {5}},\ \bibinfo {eid} {044005} (\bibinfo {year} {2019})},\ \Eprint {https://arxiv.org/abs/1712.05420} {arXiv:1712.05420 [astro-ph.IM]} \BibitemShut {NoStop}%
\bibitem [{\citenamefont {Skilling}(2006)}]{skilling06_nestedsampling}%
  \BibitemOpen
  \bibfield  {author} {\bibinfo {author} {\bibfnamefont {J.}~\bibnamefont {Skilling}},\ }\href {https://doi.org/10.1214/06-BA127} {\bibfield  {journal} {\bibinfo  {journal} {Bayesian Analysis}\ }\textbf {\bibinfo {volume} {1}},\ \bibinfo {pages} {833 } (\bibinfo {year} {2006})}\BibitemShut {NoStop}%
\bibitem [{\citenamefont {Bradbury}\ \emph {et~al.}(2024)\citenamefont {Bradbury}, \citenamefont {Frostig}, \citenamefont {Hawkins}, \citenamefont {Johnson}, \citenamefont {Leary}, \citenamefont {Maclaurin}, \citenamefont {Necula}, \citenamefont {Paszke}, \citenamefont {Vander{P}las}, \citenamefont {Wanderman-{M}ilne},\ and\ \citenamefont {Zhang}}]{jaxgithub}%
  \BibitemOpen
  \bibfield  {author} {\bibinfo {author} {\bibfnamefont {J.}~\bibnamefont {Bradbury}}, \bibinfo {author} {\bibfnamefont {R.}~\bibnamefont {Frostig}}, \bibinfo {author} {\bibfnamefont {P.}~\bibnamefont {Hawkins}}, \emph {et~al.},\ }\href {http://github.com/google/jax} {\bibinfo {title} {{JAX}: composable transformations of {P}ython+{N}um{P}y programs}} (\bibinfo {year} {2024})\BibitemShut {NoStop}%
\bibitem [{\citenamefont {{Albert}}(2020)}]{albert20_jaxns1}%
  \BibitemOpen
  \bibfield  {author} {\bibinfo {author} {\bibfnamefont {J.~G.}\ \bibnamefont {{Albert}}},\ }\href {https://doi.org/10.48550/arXiv.2012.15286} {\bibfield  {journal} {\bibinfo  {journal} {arXiv e-prints}\ ,\ \bibinfo {eid} {arXiv:2012.15286}} (\bibinfo {year} {2020})},\ \Eprint {https://arxiv.org/abs/2012.15286} {arXiv:2012.15286 [astro-ph.IM]} \BibitemShut {NoStop}%
\bibitem [{\citenamefont {{Albert}}(2023)}]{albert23_jaxns2}%
  \BibitemOpen
  \bibfield  {author} {\bibinfo {author} {\bibfnamefont {J.~G.}\ \bibnamefont {{Albert}}},\ }\href {https://doi.org/10.48550/arXiv.2312.11330} {\bibfield  {journal} {\bibinfo  {journal} {arXiv e-prints}\ ,\ \bibinfo {eid} {arXiv:2312.11330}} (\bibinfo {year} {2023})},\ \Eprint {https://arxiv.org/abs/2312.11330} {arXiv:2312.11330 [astro-ph.IM]} \BibitemShut {NoStop}%
\bibitem [{\citenamefont {Kass}\ and\ \citenamefont {Raftery}(1995)}]{kass+95_bayesfactortable}%
  \BibitemOpen
  \bibfield  {author} {\bibinfo {author} {\bibfnamefont {R.~E.}\ \bibnamefont {Kass}}\ and\ \bibinfo {author} {\bibfnamefont {A.~E.}\ \bibnamefont {Raftery}},\ }\href {https://doi.org/10.1080/01621459.1995.10476572} {\bibfield  {journal} {\bibinfo  {journal} {Journal of the American Statistical Association}\ }\textbf {\bibinfo {volume} {90}},\ \bibinfo {pages} {773} (\bibinfo {year} {1995})}\BibitemShut {NoStop}%
\bibitem [{\citenamefont {{Llorente}}\ \emph {et~al.}(2022)\citenamefont {{Llorente}}, \citenamefont {{Martino}}, \citenamefont {{Curbelo}}, \citenamefont {{Lopez-Santiago}},\ and\ \citenamefont {{Delgado}}}]{llorente+22_priorproblems}%
  \BibitemOpen
  \bibfield  {author} {\bibinfo {author} {\bibfnamefont {F.}~\bibnamefont {{Llorente}}}, \bibinfo {author} {\bibfnamefont {L.}~\bibnamefont {{Martino}}}, \bibinfo {author} {\bibfnamefont {E.}~\bibnamefont {{Curbelo}}}, \emph {et~al.},\ }\href {https://doi.org/10.48550/arXiv.2206.05210} {\bibfield  {journal} {\bibinfo  {journal} {arXiv e-prints}\ ,\ \bibinfo {eid} {arXiv:2206.05210}} (\bibinfo {year} {2022})},\ \Eprint {https://arxiv.org/abs/2206.05210} {arXiv:2206.05210 [stat.ME]} \BibitemShut {NoStop}%
\bibitem [{\citenamefont {Petrone}\ \emph {et~al.}(2014)\citenamefont {Petrone}, \citenamefont {Rizzelli}, \citenamefont {Rousseau},\ and\ \citenamefont {Scricciolo}}]{petrone+14_empiricalbayes}%
  \BibitemOpen
  \bibfield  {author} {\bibinfo {author} {\bibfnamefont {S.}~\bibnamefont {Petrone}}, \bibinfo {author} {\bibfnamefont {S.}~\bibnamefont {Rizzelli}}, \bibinfo {author} {\bibfnamefont {J.}~\bibnamefont {Rousseau}},\ and\ \bibinfo {author} {\bibfnamefont {C.}~\bibnamefont {Scricciolo}},\ }\href {https://doi.org/10.1007/s40300-014-0044-1} {\bibfield  {journal} {\bibinfo  {journal} {METRON}\ }\textbf {\bibinfo {volume} {72}},\ \bibinfo {pages} {201} (\bibinfo {year} {2014})}\BibitemShut {NoStop}%
\bibitem [{\citenamefont {{Arzoumanian}}\ \emph {et~al.}(2020)\citenamefont {{Arzoumanian}}, \citenamefont {{Baker}}, \citenamefont {{Brazier}}, \citenamefont {{Brook}}, \citenamefont {{Burke-Spolaor}}, \citenamefont {{B{\'e}csy}}, \citenamefont {{Charisi}}, \citenamefont {{Chatterjee}}, \citenamefont {{Cordes}}, \citenamefont {{Cornish}}, \citenamefont {{Crawford}}, \citenamefont {{Cromartie}}, \citenamefont {{Crowter}}, \citenamefont {{Decesar}}, \citenamefont {{Demorest}}, \citenamefont {{Dolch}}, \citenamefont {{Elliott}}, \citenamefont {{Ellis}}, \citenamefont {{Ferdman}}, \citenamefont {{Ferrara}}, \citenamefont {{Fonseca}}, \citenamefont {{Garver-Daniels}}, \citenamefont {{Gentile}}, \citenamefont {{Good}}, \citenamefont {{Hazboun}}, \citenamefont {{Islo}}, \citenamefont {{Jennings}}, \citenamefont {{Jones}}, \citenamefont {{Kaiser}}, \citenamefont {{Kaplan}}, \citenamefont {{Kelley}}, \citenamefont {{Key}}, \citenamefont {{Lam}}, \citenamefont {{Lazio}}, \citenamefont {{Levin}}, \citenamefont {{Luo}},
  \citenamefont {{Lynch}}, \citenamefont {{Madison}}, \citenamefont {{McLaughlin}}, \citenamefont {{Mingarelli}}, \citenamefont {{Ng}}, \citenamefont {{Nice}}, \citenamefont {{Pennucci}}, \citenamefont {{Pol}}, \citenamefont {{Ransom}}, \citenamefont {{Ray}}, \citenamefont {{Shapiro-Albert}}, \citenamefont {{Siemens}}, \citenamefont {{Simon}}, \citenamefont {{Spiewak}}, \citenamefont {{Stairs}}, \citenamefont {{Stinebring}}, \citenamefont {{Stovall}}, \citenamefont {{Swiggum}}, \citenamefont {{Taylor}}, \citenamefont {{Vallisneri}}, \citenamefont {{Vigeland}}, \citenamefont {{Witt}}, \citenamefont {{Zhu}},\ and\ \citenamefont {{NANOGrav Collaboration}}}]{arzoumanian+20}%
  \BibitemOpen
  \bibfield  {author} {\bibinfo {author} {\bibfnamefont {Z.}~\bibnamefont {{Arzoumanian}}}, \bibinfo {author} {\bibfnamefont {P.~T.}\ \bibnamefont {{Baker}}}, \bibinfo {author} {\bibfnamefont {A.}~\bibnamefont {{Brazier}}}, \emph {et~al.},\ }\href {https://doi.org/10.3847/1538-4357/ababa1} {\bibfield  {journal} {\bibinfo  {journal} {\apj}\ }\textbf {\bibinfo {volume} {900}},\ \bibinfo {eid} {102} (\bibinfo {year} {2020})},\ \Eprint {https://arxiv.org/abs/2005.07123} {arXiv:2005.07123 [astro-ph.GA]} \BibitemShut {NoStop}%
\bibitem [{\citenamefont {{Antil}}\ \emph {et~al.}(2012)\citenamefont {{Antil}}, \citenamefont {{Field}}, \citenamefont {{Herrmann}}, \citenamefont {{Nochetto}},\ and\ \citenamefont {{Tiglio}}}]{antil_2012_roq}%
  \BibitemOpen
  \bibfield  {author} {\bibinfo {author} {\bibfnamefont {H.}~\bibnamefont {{Antil}}}, \bibinfo {author} {\bibfnamefont {S.~E.}\ \bibnamefont {{Field}}}, \bibinfo {author} {\bibfnamefont {F.}~\bibnamefont {{Herrmann}}}, \emph {et~al.},\ }\href {https://doi.org/10.48550/arXiv.1210.0577} {\bibfield  {journal} {\bibinfo  {journal} {arXiv e-prints}\ ,\ \bibinfo {eid} {arXiv:1210.0577}} (\bibinfo {year} {2012})},\ \Eprint {https://arxiv.org/abs/1210.0577} {arXiv:1210.0577 [cs.NA]} \BibitemShut {NoStop}%
\bibitem [{\citenamefont {Leslie}\ \emph {et~al.}(2021)\citenamefont {Leslie}, \citenamefont {Dai},\ and\ \citenamefont {Pratten}}]{leslie_2021_ModebymodeRelativeBinning}%
  \BibitemOpen
  \bibfield  {author} {\bibinfo {author} {\bibfnamefont {N.}~\bibnamefont {Leslie}}, \bibinfo {author} {\bibfnamefont {L.}~\bibnamefont {Dai}},\ and\ \bibinfo {author} {\bibfnamefont {G.}~\bibnamefont {Pratten}},\ }\href {https://doi.org/10.1103/PhysRevD.104.123030} {\bibfield  {journal} {\bibinfo  {journal} {Physical Review D}\ }\textbf {\bibinfo {volume} {104}},\ \bibinfo {pages} {123030} (\bibinfo {year} {2021})}\BibitemShut {NoStop}%
\bibitem [{\citenamefont {Cornish}(2021)}]{cornish_2021_HeterodynedLikelihoodRapid}%
  \BibitemOpen
  \bibfield  {author} {\bibinfo {author} {\bibfnamefont {N.~J.}\ \bibnamefont {Cornish}},\ }\href {https://doi.org/10.1103/PhysRevD.104.104054} {\bibfield  {journal} {\bibinfo  {journal} {Physical Review D}\ }\textbf {\bibinfo {volume} {104}},\ \bibinfo {pages} {104054} (\bibinfo {year} {2021})}\BibitemShut {NoStop}%
\end{thebibliography}%

\end{document}